\documentclass[reprint,aps,amsmath,amssymb,floatfix,prb,superscriptaddress]{revtex4-2}
\usepackage{graphicx}
\usepackage{dcolumn}
\usepackage{bm}
\usepackage{xr-hyper}
\usepackage{hyperref}

	\hypersetup{
	   	bookmarks		= false,		
		unicode			= false,	
		pdfborder			= {0 0 0}	
		pdftoolbar			= false,		
		pdfmenubar		= true,		
		pdffitwindow		= false,     	
		pdfstartview		= {FitH},    	
		pdfnewwindow	= true,      	
		colorlinks			= true,		
		linkcolor			= blue,		
		citecolor			= blue,		
		filecolor			= blue,		
		urlcolor			= blue,		
		linktocpage		= true,		
		pdftitle				= {},
		pdfauthor			= {},
		pdfsubject			= {},
		pdfcreator			= {MikTeX},
		pdfproducer		= {},
		pdfkeywords		= {},
}

\begin{document}

\preprint{APS/123-QED}

\title{Interplay of crystal thickness and in-plane anisotropy and evolution of quasi-one dimensional electronic character in ReSe$_{2}$}

\author{Lewis S. Hart}
\affiliation{Centre for Nanoscience and Nanotechnology and Department of Physics, University of Bath, Bath BA2 7AY, United Kingdom}
\author{Surani M. Gunasekera}
\affiliation{Centre for Nanoscience and Nanotechnology and Department of Physics, University of Bath, Bath BA2 7AY, United Kingdom}
\author{Marcin Mucha-Kruczy\'{n}ski}
\affiliation{Centre for Nanoscience and Nanotechnology and Department of Physics, University of Bath, Bath BA2 7AY, United Kingdom}
\author{James L. Webb}
\affiliation{Center for Macroscopic Quantum States (bigQ), Department of Physics, Technical University of Denmark, Kgs. Lyngby, Denmark}
\author{Jos\'{e} Avila}
\affiliation{Synchrotron SOLEIL, Saint Aubin and Universit\'{e} Paris-Saclay, BP 48 91192 Gif-sur-Yvette, France}
\author{Mar\'{\i}a C. Asensio}
\affiliation{Madrid Institute of Materials Science (ICMM), Spanish Scientific Research Council (CSIC) and CSIC associated unit MATIN\'{E}E (ICMM and Institute of Materials Science of Valencia University, ICMUV) Cantoblanco, E-28049 Madrid, Spain}
\author{Daniel Wolverson}
\email{d.wolverson@bath.ac.uk}
\affiliation{Centre for Nanoscience and Nanotechnology and Department of Physics, University of Bath, Bath BA2 7AY, United Kingdom}


\begin{abstract}
We study the valence band structure of ReSe$_{2}$ crystals with varying thickness down to a single layer using nanoscale angle-resolved photoemission spectroscopy and density functional theory. The width of the top valence band in the direction perpendicular to the rhenium chains decreases with decreasing number of layers, from $\sim 200$~meV for the bulk to $\sim 80$~meV for monolayer. This demonstrates increase of in-plane anisotropy induced by changes in the interlayer coupling and suggests progressively more one-dimensional character of electronic states in few-layer rhenium dichalcogenides.
\end{abstract}

\maketitle

\section{Introduction}

The transition metal dichalcogenides (TMDs) present a fascinating diversity of behaviour including: thickness-driven indirect-to-direct semiconducting band gap transitions; metals; magnetic phases; charge density wave materials; superconductors, and materials with topological surface states \cite{Bahramy2018}. They are readily combined in vertical heterostructures with atomically precise interfaces and offer unprecedented degrees of built-in control over their optical \cite{Mak2016} and electronic \cite{Wang2012} properties via novel means such as variation of the heterostructure twist angle \cite{Nayak2017} and more conventional methods, such as strain \cite{He2016} or electrostatic gating \cite{Li2016}.

Within the TMD family, ReSe$_{2}$ and ReS$_{2}$ are distinguished by in-plane anisotropy due to a Jahn-Teller distortion which breaks the $120^{\circ}$  in-plane rotational symmetry common to two-dimensional crystals and leads to the formation of rhenium chains \cite{LAMFERS199634}. It also causes out-of-plane buckling of chalcogens which results in lack of atomic registry between consecutive layers and contributes to weakening of the interlayer interaction. The in-plane anisotropy and weak interlayer coupling impact many of the properties of rhenium dichalcogenides, from charge transport and optical absorption \cite{Rahman2017} to high catalytic activity as shown in solar water splitting \cite{Wang2016,Aliaga2017,Zhao2018}.

One of the most intriguing fundamental questions associated with layered materials is the understanding of the changes occurring as they are thinned down from three-dimensional bulk to single, two-dimensional atomic planes. In many materials such as 2H semiconducting TMDs \cite{Splendiani2010,Jin2013,Zhang2014B}, post-transition-metal monochalcogenides \cite{Jung2015,PhysRevB.96.035407,hamer2019} and black phosphorus \cite{yuan2015}, parts of the valence band are formed by orbitals extending significantly in the out-of-plane direction (e.g., $p_{z}$ or $d_{z^{2}}$) which are strongly affected by the presence of neighbouring layers and interlayer coupling. As a result, the shape of the valence band depends sensitively on the number of layers in the crystal leading to, for example, indirect-to-direct band gap transitions \cite{Splendiani2010,Jin2013,Zhang2014B}. In contrast, in rhenium TMDs, it is rhenium $d_{xy}$ and $d_{x^{2}-y^{2}}$ rather than $d_{z^{2}}$ or chalcogen $p_{z}$ orbitals that form the top of the valence band \cite{Kertesz1984, Choi2020}. As these are the orbitals participating in the Jahn-Teller distortion, this poses the question of whether the in-plane anisotropy of the valence band could be influenced by tuning the interlayer interaction.

Here, we use nanoscale angle-resolved photoemission spectroscopy (nano-ARPES) supported by density functional theory calculations to study the interplay between the in-plane anisotropy and crystal thickness in the electronic band structure of ReSe$_{2}$, from bulk down to the monolayer limit. We show that, as previously argued for this compound \cite{Choi2020}, the position of the valence band maximum does not change with the number of layers. However, we observe a decrease in the width of the top valence band in the direction perpendicular to the rhenium chains with decreasing thickness, indicating the increasing decoupling of rhenium chains and the growing one-dimensional character of the electronic states.

\section{Methods}
Sample preparation, characterisation details, and all relevant raw data are given in Sections S1, S2 and S3, respectively, of the Supplemental Material (SM) \cite{SM}. Nano-ARPES data were obtained at the ANTARES beamline of the SOLEIL synchrotron, Paris, which is equipped with a zone plate allowing a spot size of 120~nm, an angular resolution of $\sim 0.2^\circ$ and an energy resolution of $\sim 10$~meV. The high spatial resolution allows one to map the photoemission signal as a function of position with sub-micron resolution and thus to select flakes of different thickness. In this work, photon energies of 100~eV were used since the performance of the zone plate is optimal at this energy; the photoionisation cross-sections of Re $5d$ and Se $4p$ at 100~eV are comparable and are of order 0.1 to 0.2 Mbarn \cite{Yeh1985}. For the bulk flakes, this photon energy probes states with in-plane momenta lying in a plane near the $Z$ point at the top of the Brillouin zone \cite{Hart2017}. The sample cleaning process (see SM) resulted in layers which showed atomically clean and flat regions between the bubbles of trapped contaminants; no drift due to charging effects was observed. Raman spectroscopy, core-level nanoscale scanning photoemission microscopy (nano-XPS) and atomic force microscopy (AFM) were used as characterization tools.

Valence band structures were calculated using first principles plane wave, pseudopotential-based methods within density functional theory (DFT) as implemented in the Quantum Espresso package \cite{Giannozzi2009}. Both local density (LDA) and generalised gradient (GGA) approximations for the exchange-correlation functionals were used with the projector augmented wave (PAW) method \cite{Blochl1994,Kresse1999} being used in all cases. The valence of Re was taken as 15 (configuration $5s^2 5p^6 5d^5 6s^2$). $8\times8\times1$ Monkhorst-Pack \cite{Monkhorst1976} $k$-point meshes were used for monolayers, bilayers and trilayers (12, 24 and 36 atoms per unit cell respectively), and at least $6\times6\times6$ were used for the bulk (12 atoms). Kinetic energy cutoffs were, typically, $>800$~eV. More computational details are given elsewhere \cite{Wolverson2014,Gunasekera2018} where results using scalar- and fully-relativisitic pseudopotentials are compared. Here, scalar-relativistic pseudopotentials were chosen in order to obtain wavefunction projections onto atomic states classified only by orbital angular momentum (i.e, without spin). Comparisons of calculated 1L bandstructures with and without spin-orbit coupling (SOC) have already been reported and the features of interest here are still present when SOC is included \cite{Wolverson2014, Zhao2015, Zhong2015}.

For the band structure calculations, the unit cell parameters and atomic positions were taken from experimental data \cite{LAMFERS199634} and were relaxed to obtain residual forces generally below 0.01 eV${\AA}^{-1}$. All unit cell parameters were allowed to vary, including the separation between neighboring layers in the bilayer, trilayer and bulk cases. Representative calculations are shown in Section S4 of the SM for monolayer, bilayer, trilayer and bulk band structures in the $\Gamma-M$ direction \cite{SM}. Our earlier calculations \cite{Hart2017} used the LDA with fully-relativistic pseudopotentials and gave good agreement with ARPES and crystallographic data for bulk ReSe$_2$ despite not taking inter-layer van der Waals interactions into account explicitly; in Section S4 of the SM, we also show calculations of the band structure of a bilayer in the $\Gamma-M$ direction, comparing GGA results without van der Waals corrections and with two Grimme-type dispersion corrections, GGA+D2 and GGA+D3 \cite{Grimme2010}.

\begin{figure}[t]
  \includegraphics[width=1.00\linewidth]{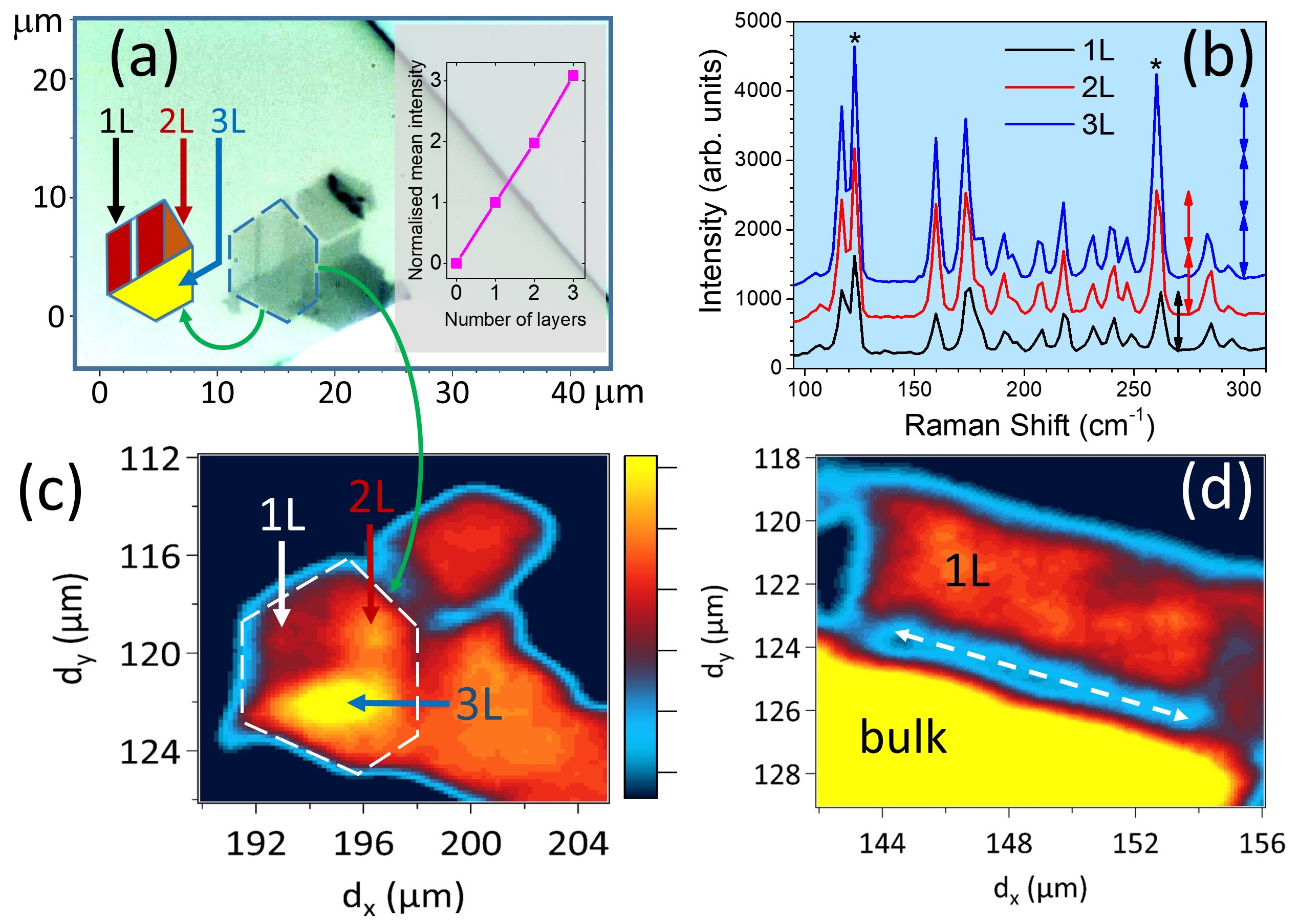}
  \caption{Structure of the samples investigated. (a) Optical microscopy image of a few-layer ReSe$_2$ flake on an HOPG platform with the region of interest bounded by the dashed blue border; the sketch of this (on the left) identifies mono-, bi- and tri-layer areas (marked 1L, 2L 3L respectively) and the inset shows the dependence of the Raman intensity (averaged over two prominent peaks) on the number of layers; (b) micro-Raman spectra of the regions of the sample shown in (a), with asterisks indicating the two peaks used to calculate the mean intensity. A vertical shift has been applied for clarity; (c) nano-XPS map of the intensity of the Re 4$f$ core levels (binding energy 40-44~eV) of the sample shown in (a) with the same region of interest indicated by the white dashed line; (d) nano-XPS intensity map of Re 4$f$ core levels for a second sample with large monolayer and bulk-like regions; the dashed arrow shows a cleavage direction of this flake. }
  \label{fgr:samples}
\end{figure}

\section{Results}\label{results}

\subsection{Characterization of atomically thin ReSe$_2$}\label{character}

Studying the dependence of the physical properties of layered materials as a function of their thickness requires samples with well-established layer numbers down to one monolayer. This task is more challenging than usual for the rhenium-based TMDs since, to enable comparisons between different flakes, their crystallographic orientations must be established including, possibly, their vertical orientation \cite{Hart2016, McCreary2017} as turning a flake upside-down is not a symmetry operation for ReSe$_2$. We have overcome this problem by exfoliating large flakes which contain regions of different thicknesses as shown in Fig.~\ref{fgr:samples} (see also Figs.~S1 and S2). Fig.~\ref{fgr:samples}(a) shows an optical micrograph of a sample whose leftmost flake contains monolayer (1L), bilayer (2L) and trilayer (3L) regions, identified on the sketch to the left of the flake. All the regions derive from the same parent crystal and have preserved the same orientation as implied by the visible cleavage edges \cite{Liu2015} at $\sim 60^\circ$ and $\sim 120^\circ$ and as confirmed rigorously by the Raman spectra of Fig.~\ref{fgr:samples}(b). Any significant rotation of one region with respect to the others would result in changes in the relative intensities of the different peaks within the spectrum due to the pronounced in-plane anisotropy of ReSe$_2$ \cite{Wolverson2014,Chenet2015,Hart2016,McCreary2017}. The orientation of the flakes is supported by a reciprocal space map of photoemission intensity for the monolayer, shown in Fig.~S3, which was recorded near the energy of the valence band maximum and displays the characteristic anisotropy already known from bulk ReSe$_2$. We confirmed layer numbers by four independent means: Raman spectroscopy, scanning photoemission microscopy, AFM and nano-ARPES (AFM images and line scans for all samples are shown in Figs.~S1 and S2). AFM step height measurements gave a monolayer thickness of $d\sim$0.60 to 0.67~nm, comparable to the interlayer lattice parameter $c=0.6702$~nm \cite{jariwala2016,LAMFERS199634} (note, the crystallographic $c$ axis in ReSe$_2$ is tilted away from the normal to the layer). The thicknesses of the 2L and 3L regions were likewise confirmed by AFM (Fig.~S2). Moreover, the Raman intensity from these regions scales linearly with the number of layers. This is best seen by looking at two Raman bands marked with asterisks in Fig.~\ref{fgr:samples}(b) which dominate the spectrum and do not overlap strongly with other spectral features -- vertical arrows next to the peaks clearly indicate the scaling. We also fit and average areas of these peaks following subtraction of a small background and present the results in the inset of panel (a). Since the silicon substrate does not have a thick oxide layer, and because the ReSe$_2$ layers were placed on a highly oriented pyrolytic graphite (HOPG) flake on the substrate, there is no modulation of the Raman intensity due to interference effects \cite{Blake2007,Wolverson2014}. The shifts in zone centre phonon frequencies with the number of layers are smaller for ReSe$_2$ and ReS$_2$ than for most TMDs \cite{Tongay2014,Wolverson2014} so that, here, interference-free Raman intensities are more reliable than Raman peak shifts for determination of layer thicknesses. However, the measured Raman peak positions support our thickness assignments: the bulk modes at 283.6 and 260.5~cm$^{-1}$ show a small shift to 285.2 and 263.0~cm$^{-1}$ in the monolayer sample, in agreement with previous experimental and computational results \cite{Zhao2015}.

\begin{figure*}[t]
  \includegraphics[width=\textwidth]{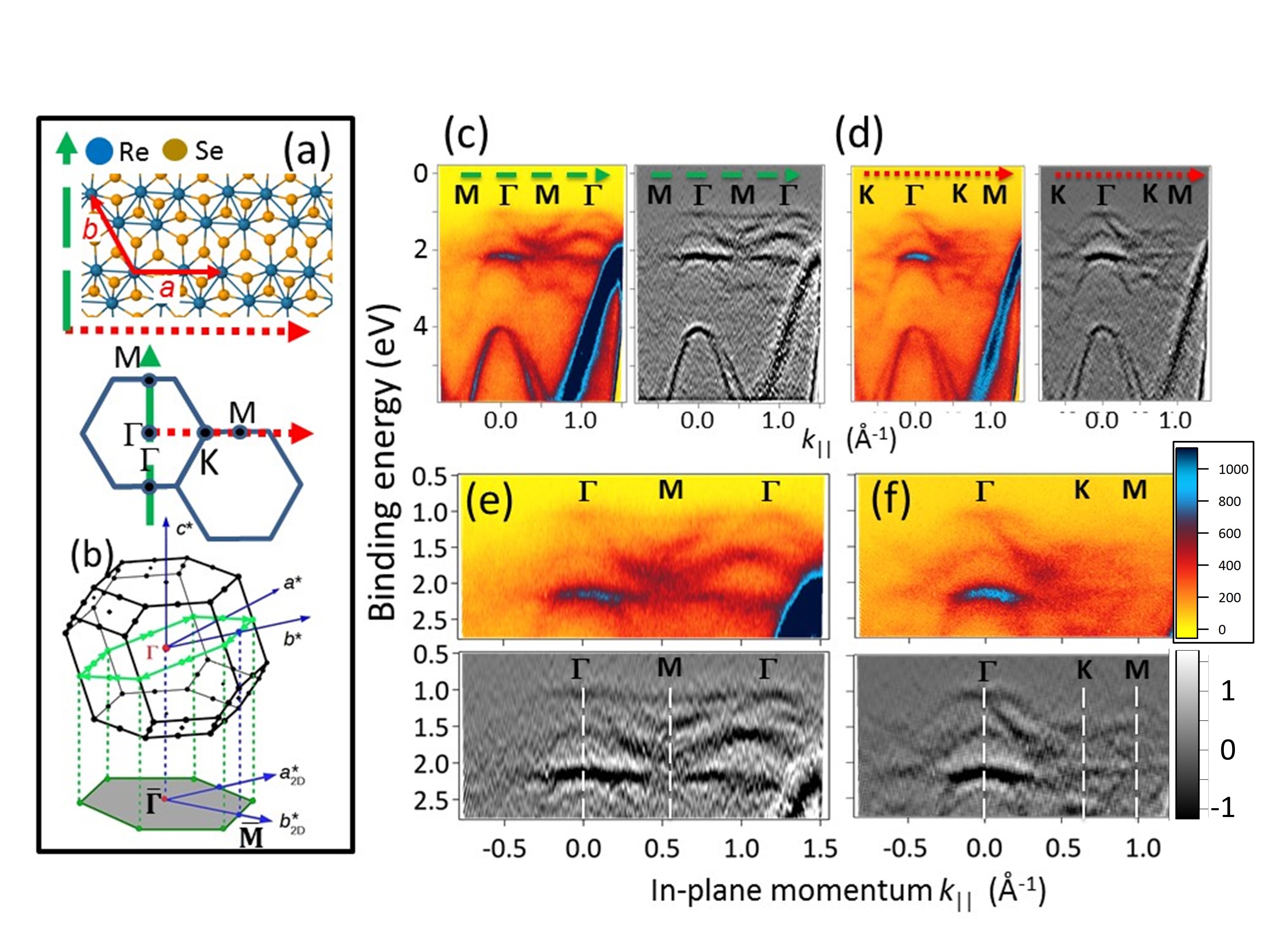}
  \caption{Valence bands of monolayer ReSe$_2$.
  (a) Crystal structure (top) and reciprocal lattice of monolayer ReSe$_2$. Solid red arrows indicate in-plane lattice vectors $a$ and $b$; the rhenium `chains' are oriented along $a$. In-plane directions $\Gamma-K$ (red dotted arrow) and $\Gamma-M$ (green dashed arrow) are displayed in relation to the top view of the real space layer plane.
  (b) First Brillouin zone of bulk ReSe$_2$ with reciprocal lattice vectors $a^*$, $b^*$ and $c^*$ and their projections onto the 2D layer plane; $\Gamma$ and $Z$ points are also shown.
  (c) Left: photoemission intensity (false colour) as a function of binding energy and in-plane wavevector for the direction $\Gamma-M$ indicated by the green dashed arrow defined in (a). The intense dispersive bands at binding energies of 2~eV and greater originate from the graphite substrate. Right: second derivatives of the
  same data.
  (d) Left: photoemission intensity as in (a) but for the in-plane wavevector in the direction $\Gamma-K$ which is close to orthogonal to the direction of (a) and which is indicated by the red dotted arrow in (a). Right: second derivatives of the same data.
  (e) and (f) Expanded views of the raw and second-derivative data of (c) and (d) respectively near the valence band edge.  }
  \label{fgr:monolayer}
\end{figure*}

Figs.~\ref{fgr:samples}(c) and (d) show maps of the photoemission intensity from the Re 4$f$ core levels (binding energy 40-44~eV) from the sample of Fig.~\ref{fgr:samples}(a) and one other sample containing monolayer and bulk regions with the same orientation. These maps were recorded primarily to locate the different regions for the ARPES experiments. Interestingly, the contrast in photoelectron count from 1L to 3L regions reflects the layer thicknesses. The core level photoemission intensity need not necessarily depend linearly on the number of ReSe$_2$ layers, but in principle it could be determined quantitatively by using X-ray photoelectron spectroscopy tools \cite{Smekal2005}. As Figs.~\ref{fgr:samples}(c) shows, discrimination between regions differing by one layer is possible and this mapping correlates with the Raman data. Imaging based on possible shifts in the \emph{energy} of the Re 4$f$ core levels \cite{Ho1999}, on the other hand, did not produce any contrast between the different regions.

\subsection{ARPES results from atomically thin ReSe$_2$}\label{1L-ARPES}

Because ReSe$_{2}$ possesses only inversion symmetry, the Brillouin zone of the monolayer is a distorted regular hexagon and the $K$ and $M$ points form three non-equivalent pairs ($K_{1},K_{2},K_{3}$, $M_{1},M_{2},M_{3}$ and their inversion equivalents). In this work, we consider only dispersions in two key directions: along, and normal to, the Re chains, for simplicity labeled $\Gamma-K$ and $\Gamma-M$ respectively; see Fig.~\ref{fgr:monolayer}(a) for the indication of these directions in the real and reciprocal spaces and (b) for comparison of the bulk and two-dimensional Brillouin zones of ReSe$_{2}$ (the latter is a projection of the former along the reciprocal space $c^*$ axis). These special directions correspond to, for example, the polarization directions of the two band edge excitons in ReSe$_2$ \cite{Ho1998,Ho2004,Ho2007} and were identified as the directions of maximum and minimum dispersion in the studies of bulk material \cite{Hart2017,Eickholt2018}. We present the measured valence band dispersions for the monolayer (1L) flake shown in Fig.~\ref{fgr:samples}(d) in Fig.~\ref{fgr:monolayer}(c,e) for $\Gamma-M$ and (d,f) for $\Gamma-K$. The right-hand panels in (c) and (d) show second derivatives with respect to energy of the data in the left, and in (e) and (f) we show expanded views of the same data and its second derivatives close to the valence band maximum (within $\sim 2$~eV). The extraordinarily flat nature of the top of the valence band in the $\Gamma-M$ direction is striking and is the main result of this work (Fig.~\ref{fgr:monolayer}(c,e)). As we show later, this is very different to the bulk valence band structure. By contrast, the top valence band remains dispersive in the near-orthogonal direction $\Gamma-K$ (Fig.~\ref{fgr:monolayer}(d,f)). Because second derivatives can occasionally generate spurious features, we have examined the energy distribution curves (EDCs) which are the raw data for Fig.~\ref{fgr:monolayer}, directly. Fig.~S4 of the SM shows these EDCs in the $\Gamma-M$ and $\Gamma-K$ directions and confirms the flatness of the band along $\Gamma-M$ while the $\Gamma-K$ direction shows significant dispersion.

One more prominent feature of the 1L ReSe$_2$ valence band structure is the locally flat band centred at $\Gamma$ at about 1~eV below the valence band edge (at a binding energy of $\sim$2~eV) which yields a particularly high photoemission intensity. This feature was shown to correspond to the valence band formed predominantly by the out-of-plane Re $d_{z^{2}}$ and Se $p_{z}$ orbitals \cite{Choi2020}. Fig.~\ref{fgr:monolayer} also shows the valence bands of the HOPG substrate on which the ReSe$_2$ flakes were placed. The observation of the $\Gamma$ point of graphite (at $k_{||}=0$) usefully confirms the experimental location of the first Brillouin zone and $\Gamma$ point of the ReSe$_2$ flakes and, as a result of the relatively large unit cell of ReSe$_2$, the accessible momentum range allows us to sample almost two complete Brillouin zones in the $\Gamma-M$ direction. The data of Fig.~\ref{fgr:monolayer} show no evidence for hybridization or appreciable opening of gaps at the binding energy and momentum values where the bands of ReSe$_2$ and HOPG intersect. These results contrast with recently reported findings of strong coupling to graphene for TMDs such as WSe$_2$ \cite{Wilson2017} and MoS$_2$ \cite{Diaz2015} though, in those studies, the flakes also had graphene capping layers which the present samples did not.

\begin{figure*}[t]
  \includegraphics[width=\textwidth]{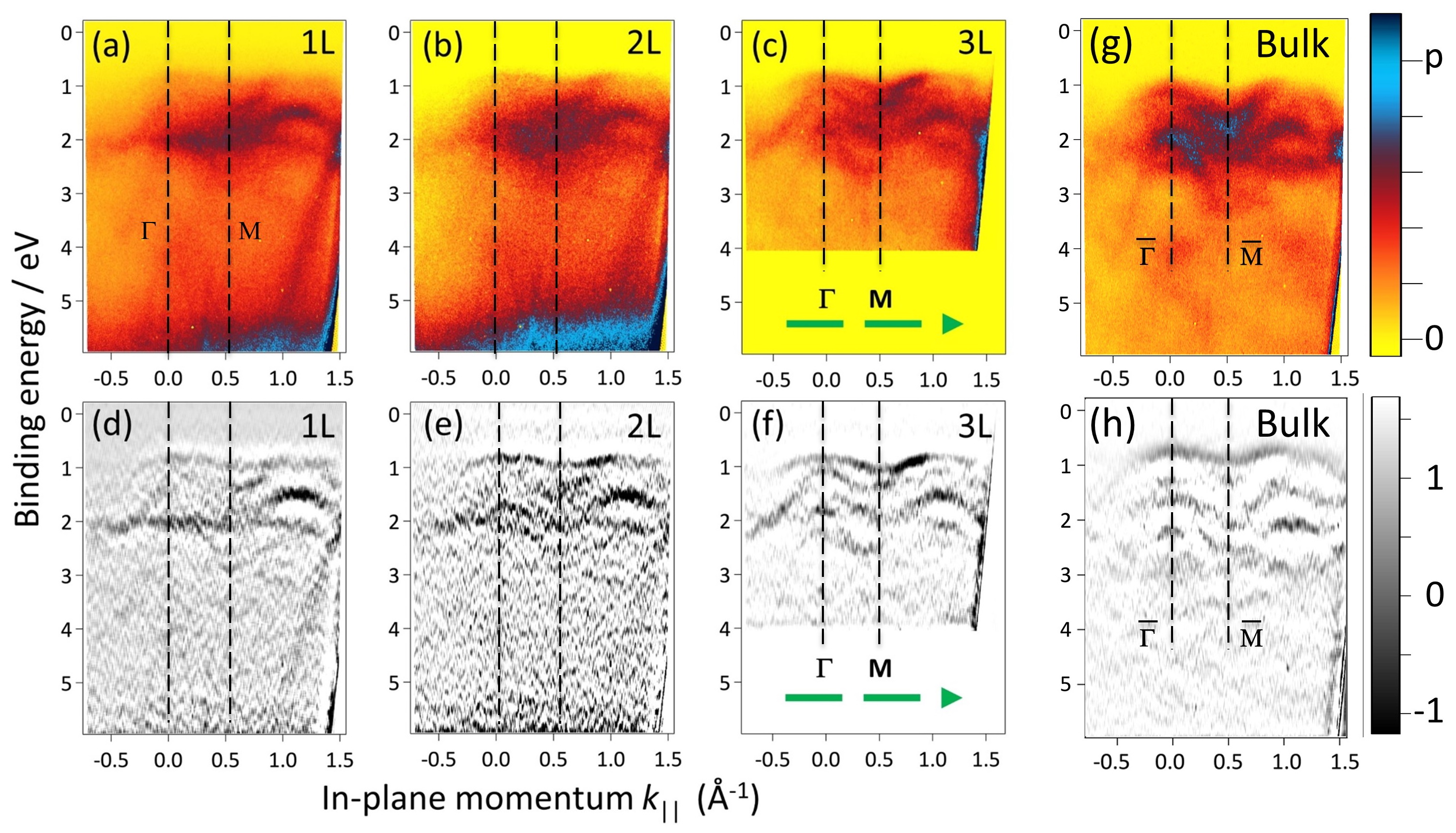}
  \caption{Comparison of the measured ReSe$_2$ valence band dispersion along $\Gamma-M$ (indicated by the green dashed arrow defined in Fig.~\ref{fgr:samples}) for mono-, bi- and tri-layer flakes (labeled 1L, 2L and 3L respectively) and bulk. (a)-(c),(g) ARPES data; (d)-(f),(h) second derivatives of the data above. Vertical dashed lines indicate the positions of the $\Gamma$ and $M$ points on each panel. The color scale applies to all the top panels and indicates counts from zero to $p$ where $p$ equals 3200, 1600 and 1600 for (a) to (c) respectively and 800 for (g).}
  \label{fgr:few-layer}
\end{figure*}

\subsection{Evolution of the valence band with the number of layers}

Following the discussion of the ARPES of the monolayer region shown in Fig.~\ref{fgr:samples}(d), we move to the flake presented in Fig.~\ref{fgr:samples}(a) and (c). Between the two samples, we can not only test the reproducibility of the 1L results of Fig.~\ref{fgr:monolayer} as well as those for the bulk, but also study the evolution of the valence band states as a function of crystal thickness. In Fig.~\ref{fgr:few-layer}, we show photointensity as a function of binding energy and momentum, (a)-(c), and its second-derivative, (d)-(f), recorded using nano-ARPES for all our few-layer flakes as well as bulk,(g) and (h), for a momentum slice in the $\Gamma-M$ direction (note, the data in (c) and (f) was not recorded over the same binding energy range as in the other panels). For this comparison, we labelled the bands of the bulk flake following the high symmetry points of the projected 2D Brillouin zone, using symbols $\bar{\Gamma}$, $\bar{M}$ and $\bar{K}$ to indicate that the vertical momentum component of the bands is not specified (at the photon energy used here (100~eV), ARPES probes states lying in an approximately planar surface passing near the $Z$ point at the top of the 3D Brillouin zone \cite{Hart2017} shown in Fig.~\ref{fgr:monolayer}(b), i.e, $\bar{\Gamma} \approx Z$). Although the data for monolayer are not as clear as those obtained with the larger sample of Fig.~\ref{fgr:monolayer}, there is a systematic trend in the band width of the uppermost valence band.

\begin{figure}[t]
  \includegraphics[width=1.00\linewidth]{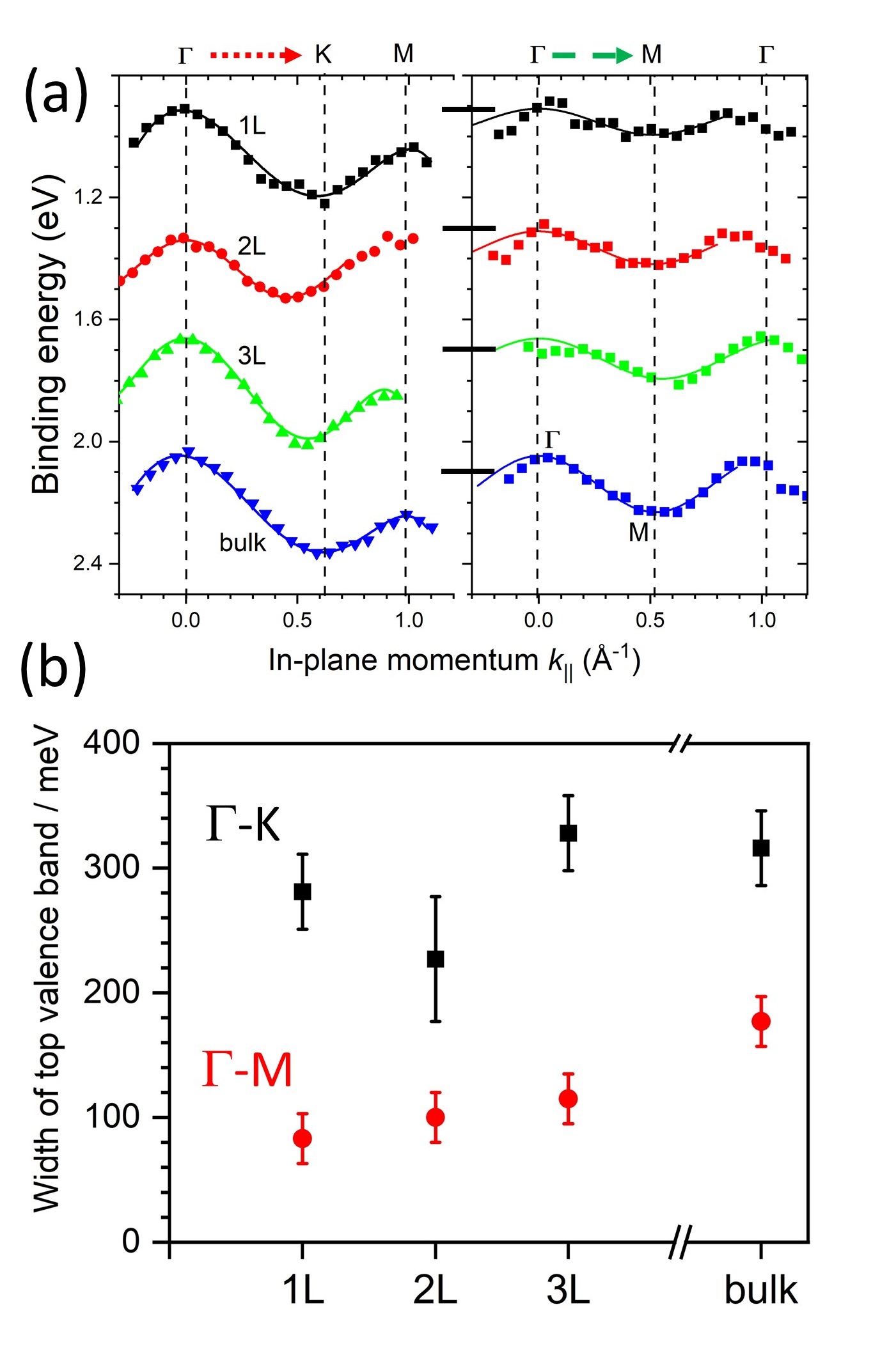}
  \caption{Comparison of the measured ReSe$_2$ valence band dispersion along $\Gamma-M$ (as indicated by the green dashed arrow defined in Fig.~\ref{fgr:samples}) for mono-, bi- and tri-layer flakes (labeled 1L, 2L and 3L respectively) and bulk; (a) band edge positions determined from an EDC fit at each wavevector, with fits to each bandedge to obtain the bandwidth. The curves for 2L, 3L and bulk samples are displaced vertically by 0.4~eV from each other for clarity (the same shifts were applied to the $\Gamma-M$ and corresponding $\Gamma-K$ curves and the 1L curves are unshifted); (b) summary of the bandwidths obtained in $\Gamma-M$ (red circles) and $\Gamma-K$ (black squares) directions for monolayer, bilayer, trilayer and bulk ReSe$_{2}$.}
  \label{fgr:summary}
\end{figure}

We use our raw data to obtain the width of that band - we extract photoemission counts as a function of binding energy (that is, the energy distribution curves, EDCs) and fit these to obtain a measure of the band edge position at each momentum along the $\Gamma-M$ and $\Gamma-K$ directions. This, in turn, allows us to identify the energies of the band extrema. The results of our procedure (for further details of the fitting process see the SM, Figs.~S5 and S6) are shown in Fig. 4(a), where we compare the extracted top valence bands for 1L, 2L, 3L and bulk. The curves for 2L, 3L and bulk are shifted vertically by steps of 0.4 eV for clarity (the same shift is applied to $\Gamma-M$ and $\Gamma-K$ in each case). The corresponding raw data are shown in the SM, Figs. S7-S9 (in Fig.~S11 we show a comparison of the raw data with the calculated band structures of bulk and monolayer samples in the $\Gamma-K$ and $\Gamma-M$ directions). A summary of the band widths obtained from fitting the EDCs is displayed in Fig.~\ref{fgr:summary}(b) which illustrates flattening of the top valence band along $\Gamma-M$ with decreasing crystal thickness while the band width along $\Gamma-K$ is comparable for three out of four investigated sample thicknesses. The error bars are estimated from the sensitivity of the measured band width to variation of the fitting parameters (principally the threshold intensity $I_{\textrm{thresh}}$ corresponding to the band edge; see Fig.~S6 for a definition of $I_{\textrm{thresh}}$). The error for the case of the bilayer in the $\Gamma-K$ direction is somewhat larger as the data is slightly poorer (see Fig.~S8); the 2L region was the smallest of the flakes (see Fig.~\ref{fgr:samples}(a)) and therefore the most difficult to measure. Note, band width values in the bulk case are photon energy-dependent due to the dispersion normal to the layers; our present values agree well with previous values measured on cleaner, cleaved bulk material \cite{Hart2017} at the same photon energy.

Our experimental findings are also supported by density functional theory (DFT) calculations. Using scalar relativistic pseudopotentials, we obtained band widths of 95 meV (1L), 150 meV (2L), 275 meV (3L) and 500 meV for bulk. A fully relativistic calculation yielded similar values of 85, 187, 217 and 436 meV, respectively. The calculated band structures along the $\Gamma-M$ direction are shown in section S4 of the SM (Fig.~S10) and the trend in their variation as a function of crystal thickness agrees well with the measurements.

At the photon energy of 100~eV used here for the nano-ARPES experiments, the recorded signal is highly surface-specific: the predicted inelastic mean free path of MoS$_2$, for example, ranges from 3.5 to 6.0~{\AA} for photon energies from 70 to 170~eV respectively, \cite{Han2012,Lince1989,Tanuma1987} comparable to the layer thickness of ReSe$_2$ ($c$-axis 6.7~{\AA}), and a simple estimate based on the universal curve \cite{Seah1979} gives 5.5~{\AA} at 100~eV. Despite this, Fig.~\ref{fgr:few-layer} and Fig.~\ref{fgr:summary} demonstrate systematic changes in the $\Gamma-M$ valence band edge as a function of the number of layers. Overall, our nano-ARPES data reveals valence band states that are clearly dependent on inter-layer hopping interactions and must be representative of the body of the flake. We do not observe electronic states confined to individual layers except in the case of the 1L sample itself. This is in agreement with studies of bulk and few-layer ReS$_2$ \cite{Biswas2017,Gehlmann2017}. Thus, although the inter-layer interactions of ReSe$_2$ are weak, very similar to ReS$_2$ \cite{Tongay2014,Zhao2015} (which shows a higher inter-layer resistivity than typical TMDs \cite{Lee2019}), they are certainly non-negligible \cite{Echeverry2018}. To compare, the interlayer coupling-induced $k_z$ dispersion of the highest valence band for $k_x=k_y=0$ in bulk ReSe$_2$ is $\sim$100~meV \cite{Hart2017} while it is $\sim$800~meV in MoSe$_2$ and $\sim$1~eV in MoS$_2$ \cite{PhysRevB.64.235305}.

Recently, the surface-sensitivity of ARPES has been applied to reveal hidden spin polarisation \cite{Zhang2014} in centrosymmetric bulk TMDs, for example, WSe$_2$ \cite{Riley2014,Bertoni2016}, MoS$_2$ \cite{Gehlmann2016}, NbSe$_2$ \cite{Bawden2016}, and PtSe$_2$ \cite{Yao2017}. In the nomenclature of spin-polarisation effects introduced by Zhang \emph{et al.} \cite{Zhang2014}, D-1 signifies a spin polarization arising from conventional bulk Dresselhaus inversion asymmetry and D-2 implies localised Dresselhaus spin polarizations compensated by their opposites under bulk inversion symmetry. The case of MoS$_2$ was considered theoretically \cite{Liu2015B}; $2H$-MoS$_2$ is a system where individual D-1 layers interact (weakly) to give D-2 behaviour in bulk. Surface-sensitive techniques such as STM or ARPES can probe the top layer of a bulk crystal and can reveal its D-1 nature. However, once again, ReX$_2$ proves to be an untypical member of the TMD family, because its structure contains a centre of inversion at the midpoint of each diamond of four Re atoms. On symmetry grounds, therefore, even a monolayer is only expected to show D-2 behaviour. The opposing spin polarisations reside on Re atoms located close to one another in the same layer, meaning that the near single-layer sensitivity of ARPES cannot resolve individual D-1 contributions. This is borne out by the lack of observed spin-orbit splittings here or in any earlier ARPES data on bulk or few-layer ReX$_2$ \cite{Choi2020,Webb2017,Biswas2017,Gehlmann2017,Hart2017,Eickholt2018}.

\subsection{In-plane anisotropy and interlayer coupling}\label{1L-vs-bulk}

\begin{figure*}[t]
  \includegraphics[width=\textwidth]{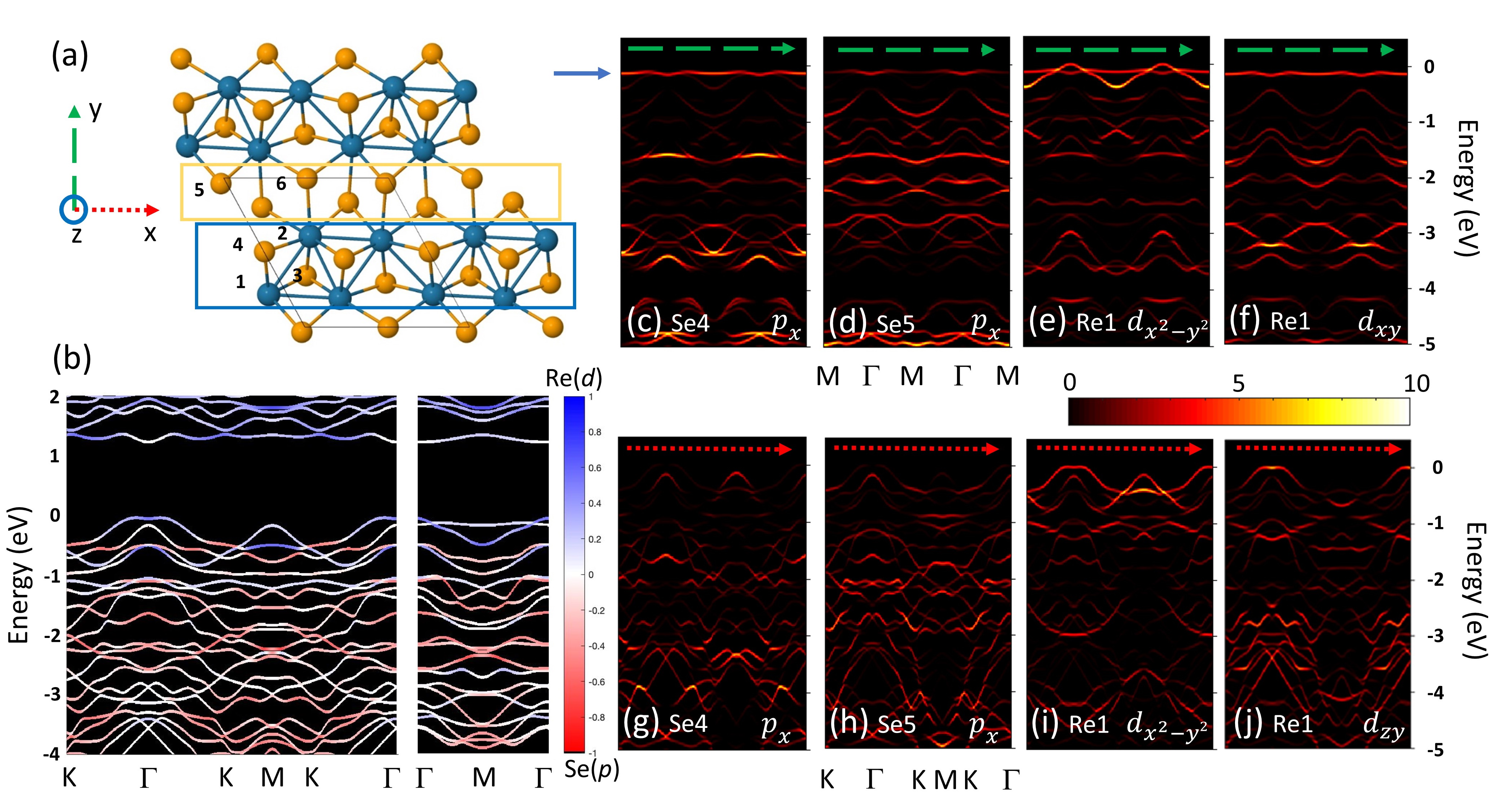}
  \caption{The valence band structure of monolayer ReSe$_2$ analysed in terms of the atomic orbitals of rhenium and selenium. (a) top view of ReSe$_2$ showing the numbering of the atoms: 1,2 are Re, 3-6 are Se. Other atoms are related to these by symmetry. Rectangular boxes show the assignment of the chalcogen atoms to two families, those between rhenium chains (upper yellow box) and those above and below the chains (lower blue box). (b) Calculated valence band dispersion of monolayer ReSe$_2$ with states projected onto Se $p$ and Re $d$ orbitals (normalised amplitudes shown using the red-blue false color scale shown on the right; white indicates an equal admixture). (c)-(j) Calculated valence band states projected onto the selected Re and Se atomic orbitals identified on each panel for the in-plane momentum directions $\Gamma-M$ and $\Gamma-K$ defined in Fig.~\ref{fgr:monolayer} (reciprocal space directions are indicated by dashed green arrows and dotted red arrows respectively at the top of each panel). The top of the valence band is defined as the zero of energy and the blue arrow by panel (c) marks the near-dispersionless state discussed in the text. The color scale in units of states per eV applying to panels (c)-(j) is shown below panels (e,f).}
  \label{fgr:projections}
\end{figure*}

In order to understand better the anisotropic change in the band width of the top valence band as a function of the number of layers, we investigated the orbital character of the valence band of monolayer and bulk ReSe$_{2}$ using density functional theory calculations and show the key results for monolayer in Fig.~\ref{fgr:projections} (the complete set of projections for the monolayer as well as bulk are presented in Section S4 of the SM). In these calculations, we have neglected the spin-orbit coupling so that the atomic wavefunctions are purely orbital angular momentum states (this is acceptable because inversion symmetry which forbids any band spin splitting is present for all crystal thicknesses, even monolayer). For this discussion, $x,y$ are defined as the directions along and normal to the Re chains respectively and $z$ is normal to the layer, as in Fig.~\ref{fgr:projections}(a).

In Fig.~\ref{fgr:projections}(b), we show the valence band dispersion of monolayer ReSe$_{2}$ along the directions $\Gamma-K$ and $\Gamma-M$ (calculated at the scalar relativistic GGA level) projected onto Se $p$ and Re $d$ orbitals with the respective amplitudes shown using red and blue false color scales and white indicating an equal admixture. In general, the Re $d$ states dominate in the conduction bands while the Se $p$ orbitals prevail on the valence band side. However, the top valence band defies this simple picture: states in the vicinity of $\Gamma$ are built mainly of the Re $d$ orbitals while those around $K$ of Se $p$ orbitals. Along $\Gamma-M$, the Re $d$-character band forming the valence band maximum around $\Gamma$ crosses the second band made of a mixture of Re $d$- and Se $p$-orbitals so that the highest energy states in this direction contain comparable transition metal and chalcogen contributions. We suggest it is this difference between the orbital composition of the top valence band in the directions along ($\Gamma-K$) and perpendicular ($\Gamma-M$) to the rhenium chains that leads to the unique coupling between in-plane anisotropy of this band and the crystal thickness. The sandwich-like nature of TMD monolayers means that chalcogen orbitals are more involved in, and hence affected by, the interlayer interaction than the rhenium orbitals. Hence, while states in the $\Gamma-M$ direction are significantly affected by the presence of other layers in thicker crystals which is reflected by the growing dispersion in Fig.~\ref{fgr:summary}(b), only a small part of states along $\Gamma-K$ responds to interlayer coupling so that the band width in this direction remains similar in flakes of different thicknesses. In a sense, this mechanism is similar to the one driving the direct-indirect band gap transition in 2H TMDs like MoS$_2$ \cite{Splendiani2010,Jin2013,Zhang2014B} where interlayer coupling causes shifts of the highest valence band in the vicinity of $\Gamma$ (made of out-of-plane S $p_z$ and Mo $d_{z^2}$ orbitals) while the states around $K$ (formed by in-plane Mo $d$-orbitals) remain unaffected. In ReSe$_{2}$ however, out-of-plane orbitals do not contribute to the highest valence band and so the distinction is rather between the atomic species than orbital symmetries.

We investigate the orbital contributions of Re and Se to the valence band in more detail in Fig.~\ref{fgr:projections}(c)-(j). We note that the four selenium sites that are not related by symmetry make different contributions to the band structure and, of these, one can identify two main types of chalcogen site: those located on the Re diamonds, labeled atoms 3 and 4 in Fig.~\ref{fgr:projections}(a), and those bridging adjacent Re chains (atoms 5 and 6) which differ markedly. Evidence for the non-equivalence of these sites is also provided by Raman spectroscopy studies of ReSe$_{2-x}$S$_x$ alloys which showed that the substitution of sulphur on the different chalcogen sites yields different formation energies and vibrational frequencies \cite{Hart2017} whilst high resolution electron microscopy suggests preferential occupation of the more stable sites by impurities \cite{Wen2017}. Scanning tunneling microscopy (STM) and spectroscopy (STS) likewise show very clearly the non-equivalence of the four selenium sites in ReSe$_2$ \cite{Hong2018}. The projections of the valence band states onto the Se $p$ orbitals, Fig.~\ref{fgr:projections}(c) and (d), show that it is the $p_x$ orbitals of Se sites 4 (and 3, see Fig.~S15) that contribute to the flat VBM in the $\Gamma-M$ direction. As seen in Fig.~\ref{fgr:projections}(d) and (h), the contribution of Se $p_x$ of sites 5 (and 6, Fig.~S15) to the valence band edge is less significant. From the coordinate system shown in Fig.~\ref{fgr:projections}(a), it is clear that the $p_x$ states of sites 3 and 4 are polarized along the Re chains and are spatially localized above and below them. The interaction between the orbitals of Se sites 3 and 4 on neighboring chains is minimal and this gives rise to a near-absence of dispersion in the $\Gamma-M$ ($y$) direction in the monolayer. The dispersion along the $\Gamma-K$ ($x$) direction is generated predominantly by Re $d_{x^2-y^2}$ and $d_{zy}$ orbitals, Fig.~\ref{fgr:projections}(i) and (j).

\section{Summary}\label{discuss}

We have demonstrated that in ReSe$_{2}$ in-plane anisotropy is quite uniquely coupled to interlayer interaction so that decreasing the number of layers in the crystal decreases the dispersion in the direction perpendicular to the rhenium chains ($\Gamma-M$ in reciprocal space). This implies increasing interchain decoupling and a growing one-dimensional character of electronic states in this material. The extremely flat valence band dispersion perpendicular to the Re chains implies that, in monolayer ReSe$_2$, hole transport should be dominated by conduction along the direction of the Re chains and thus should be extremely anisotropic. No experimental test of this has yet been carried out, though it is established that the in-plane mobility is lowest perpendicular to the Re chains in bulk-like material. An anisotropy of about a factor of two between mobilities in the $a$ and $b$ directions was reported for few-layer ReSe$_{2-x}$S$_x$ devices \cite{Liu2016a} but their reported number of layers was $\sim 5$ which, as shown above, is too thick to observe the present effects; furthermore, they observed $n$-type conductivity. Another study found an anisotropy ratio of $\sim 4$ for $n$-type conductivity in W-doped bulk material \cite{Hu2006} in agreement with a recent ARPES study \cite{Kim2019}. Several other studies on exfoliated material likewise used few-layer flakes down to 3L, or $n$-type material, and so do not provide a test of the present predictions \cite{Corbet2015, Ali2018,Zhang2016,Yang2014a,Yang2014b}.  There is only one report of measurements on $p$-type conductivity in a monolayer, where a hole mobility of 10~cm$^2$V$^{-1}$s$^{-1}$ was found, but the anisotropy of the charge transport was apparently not investigated \cite{Yang2014b}. It is therefore a high priority to investigate transport in monolayer ReSe$_2$ in suitably designed structures which would ideally allow for investigation both of the predicted anisotropy but also electrostatic gating to ensure hole transport dominates. The quasi-one dimensional nature of hole transport in ReSe$_2$ offers the possibility of a momentum-selective filter or contact to other TMD layers in either lateral or vertical heterostructures.

\begin{acknowledgments}
This work was supported by the Centre for Graphene Science of the Universities of Bath and Exeter and by the EPSRC (UK) under grants EP/G036101, EP/M022188, and EP/P004830; L.S.H. and S.M.G. are supported by the Bath/Bristol Centre for Doctoral Training in Condensed Matter Physics, EPSRC grant EP/L015544. M.M-K. acknowledges support from the University of Bath International Funding Scheme. We thank the SOLEIL synchrotron for the provision of beam time; work at SOLEIL was supported by EPSRC grant EP/P004830/1. The present collaborative research was undertaken in the context of the Associated Unit MATIN\'{E}E of the Spanish Scientific Research Council (CSIC). Computational work was performed on the University of Bath's High Performance Computing Facility. Data created during this research are freely available from the University of Bath data archive at DOI:10.15125/BATH-00521.
\end{acknowledgments}

\bibliography{ReSe2_Refs}

\end{document}


\opensupplement

\preprint{APS/123-QED}

\title{Supplemental Material:\\ Interplay of crystal thickness and in-plane anisotropy and evolution of quasi-one dimensional electronic character in ReSe$_{2}$}

\author{Lewis S. Hart}
\author{Surani M. Gunasekera}
\author{James L. Webb}
\author{Marcin Mucha-Kruczy\'{n}ski}
\affiliation{Centre for Nanoscience and Nanotechnology and Department of Physics, University of Bath, Bath BA2 7AY, United Kingdom}
\author{Jos\'{e} Avila}
\affiliation{Synchrotron SOLEIL, Saint Aubin and Universit\'{e} Paris-Saclay, BP 48 91192 Gif-sur-Yvette, France}
\author{Mar\'{\i}a C. Asensio}
\affiliation{Madrid Institute of Materials Science (ICMM), Spanish Scientific Research Council (CSIC), Cantoblanco, E-28049 Madrid, Spain}
\author{Daniel Wolverson}
\affiliation{Centre for Nanoscience and Nanotechnology and Department of Physics, University of Bath, Bath BA2 7AY, United Kingdom}
\email{d.wolverson@bath.ac.uk}


\maketitle

\tableofcontents

\newpage

\section*{List of figures}
\begin{itemize}
  \item Figure~\ref{fgr:S1}: Optical and AFM images of a monolayer flake
  \item Figure~\ref{fgr:S2}: Optical and AFM data for few-layer flakes
  \item Figure~\ref{fgr:S3}: Reciprocal space maps near the valence band maximum for monolayer ReSe$_2$
  \item Figure~\ref{fgr:S4}: Photoemission energy distribution curves for monolayer ReSe$_2$
  \item Figure~\ref{fgr:S5}: Photoemission energy distribution curves for 1L, 2L, 3L and bulk ReSe$_2$
  \item Figure~\ref{fgr:S6}: Procedure for obtaining the VB band edge from experimental data
  \item Figure~\ref{fgr:S7}: Comparison of 1L and bulk ReSe$_2$ ARPES in $\Gamma-M$ and $\Gamma-K$ directions
  \item Figure~\ref{fgr:S8}: Measured VB dispersions of few-layer ReSe$_2$
  \item Figure~\ref{fgr:S9}: Second derivatives of VB dispersions of few-layer ReSe$_2$
  \item Figure~\ref{fgr:S10}: Calculated VB dispersion in the $\Gamma-M$ direction for few-layer and bulk ReSe$_2$
  \item Figure~\ref{fgr:S11}: Comparison of calculated and experimental dispersions of ReSe$_2$
  \item Figure~\ref{fgr:S12}: Projection of calculated VB dispersion onto Re $d$-orbitals ($\Gamma-K$)
  \item Figure~\ref{fgr:S13}: Projection of calculated VB dispersion onto Re $d$-orbitals ($\Gamma-M$)
  \item Figure~\ref{fgr:S14}: Projection of calculated VB dispersion onto Se $p$-orbitals ($\Gamma-K$)
  \item Figure~\ref{fgr:S15}: Projection of calculated VB dispersion onto Se $p$-orbitals ($\Gamma-M$)
  \item Figure~\ref{fgr:S16}: Projection of calculated VB dispersion onto Se $p$-orbitals ($\Gamma-Z$)
  \item Figure~\ref{fgr:S17}: Projection of calculated VB dispersion onto Re $d$-orbitals ($\Gamma-Z$)
\end{itemize}

\clearpage

\section{Sample preparation.} ReSe$_2$ samples were prepared in air by micro-mechanical cleavage (exfoliation) from bulk single crystals grown by chemical vapour transport and supplied by HQ Graphene. The cleaved flakes were transferred to a PDMS film for solvent-free transfer to a large highly oriented pyrolytic graphite (HOPG) flake on a conducting silicon substrate \cite{Castellanos2014}. The HOPG platform (with dimensions of several 100 $\mu$m) provides a conducting link to the grounded silicon substrate to prevent the ReSe$_2$ layers charging during measurement (essential for both imaging and spectroscopy) and assists in locating the ReSe$_2$ layers in the ARPES experiment. Samples were washed with acetone and isopropanol to remove organic residues and were then annealed in argon for 5 hours at 400~$^\circ$C; annealing results in coalescence of material trapped under the layer into relatively few, large bubbles with flat regions in between, as revealed by atomic force microscopy (AFM) and as observed also for WSe$_2$ \cite{Wilson2017} (in contrast to Ref.~\cite{Wilson2017}, however, no graphene capping layer was used). Once mounted in the beamline, samples were annealed again in UHV at 400~$^\circ$C for over 12 hours.

\section{Sample characterization.}
\begin{figure}[h]
  \includegraphics[width=\linewidth]{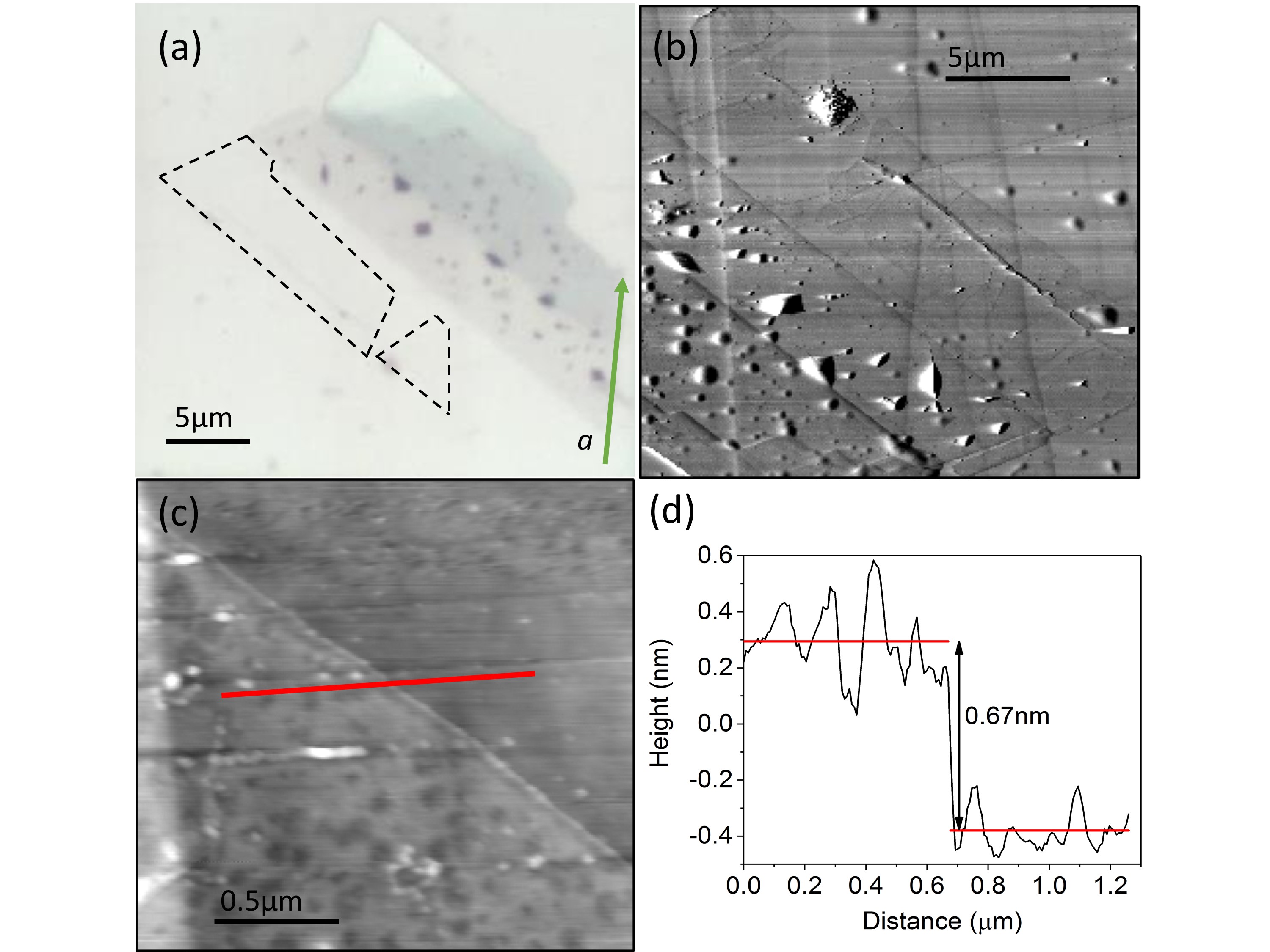}
  \caption{(a) Optical image of the ReSe$_2$ monolayer after placing on the HOPG support and annealing. The dashes show the outlines of two monolayer regions, the larger of which was used in the ARPES work; (b) an AFM phase image of that monolayer, showing bubbles of material trapped under the layer developed during annealing; (c) an AFM image of the edge of the flake showing (red solid line) the path used to make the thickness measurement in (d).}
  \label{fgr:S1}
\end{figure}

\begin{figure}
  \includegraphics[width=\linewidth]{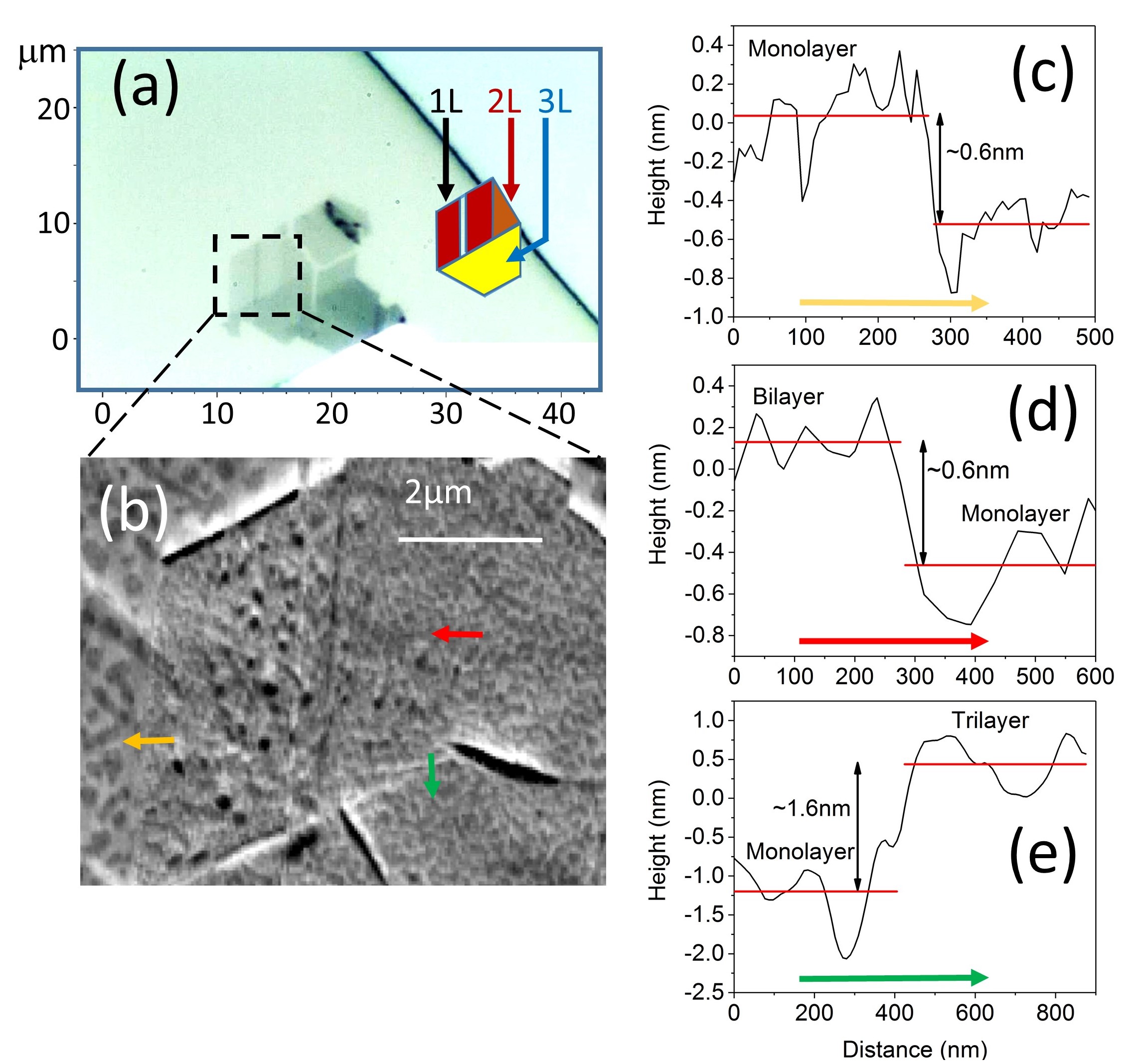}
  \caption{(a) Optical image of the 1L, 2L and 3L ReSe$_2$ flakes on the HOPG support. The sketch identifies the different regions; (b) an AFM phase image of the region indicated by the dashed box in (a). The arrows indicate the paths used for the AFM step height measurements; (c)-(e) AFM measurements of the step height (c) from HOPG to monolayer; (d) from monolayer to bilayer, and (e) from monolayer to trilayer. The AFM step height for one monolayer is consistently of order 0.6 to 0.7~nm (on the larger monolayer, we obtain 0.67~nm, Fig.~\ref{fgr:S1}) and the step in (e) is 1.6~nm, close to $2 \times 0.67 = 1.34$~nm. Given this and also the Raman data in the main text, we infer a thickness of three layers for this region.}
  \label{fgr:S2}
\end{figure}

Atomic force microscopy (AFM) line scans across step edges were used to confirm the ReSe$_2$ layer thicknesses and AFM imaging was used to assess the flatness of the layers; optical microscopy and AFM images are shown in Fig.~\ref{fgr:S1}. Layer composition and thicknesses were checked via Raman microscopy as shown in Fig.~1 of the main text; a Renishaw InVia system was used with 532~nm excitation and a $\times$~100 objective, giving a spatial resolution of better than 1 $\mu$m, adequate to select each of the 1L, 2L and 3L regions in turn. Since the ReSe$_2$ layers were placed on thick HOPG platforms and the substrate did not have a thermal SiO$_2$ layer, interference effects did not modify the intensity of the Raman spectra, which was found to depend linearly on the number of layers as discussed earlier. Finally, Re core level X-ray photoemission spectra (XPS), integrated over the binding energy range 40-44~eV, were used to image the ReSe$_2$ layers; individual XPS spectra also confirmed the bonding of Re to Se.

\clearpage

\section{Experimental data.}

\clearpage

\clearpage

\begin{figure}
  \includegraphics[scale=0.7]{figSFS.png}
  \caption{Photoemission signal (false color scale) as a function of in-plane momenta $k_x, k_y$ at (a) the energy of the valence band maximum (VBM) and (b-e) four energies below the VBM (energy values shown on the right). In this figure only, $x$ and $y$ are orthogonal, in-plane laboratory axes; elsewhere, they are defined with respect to the crystal axes as shown in Fig.~5 of the main text. The signal in each plot was integrated over an energy range of 55~meV. The plots on the right  show the same data as those on the left but with increased contrast and a smaller momentum range around $k=0$ ($\Gamma$). The first Brillouin zone superimposed on (c) indicates the directions $\Gamma-M$ and $\Gamma-K$; the flatter valence band in the $\Gamma-M$ direction is clear in the energy range of panels (b) and (c) and confirmed the crystal orientation in the experiment.}
  \label{fgr:S3}
\end{figure}
\clearpage

\begin{figure}
  \includegraphics[width=\linewidth]{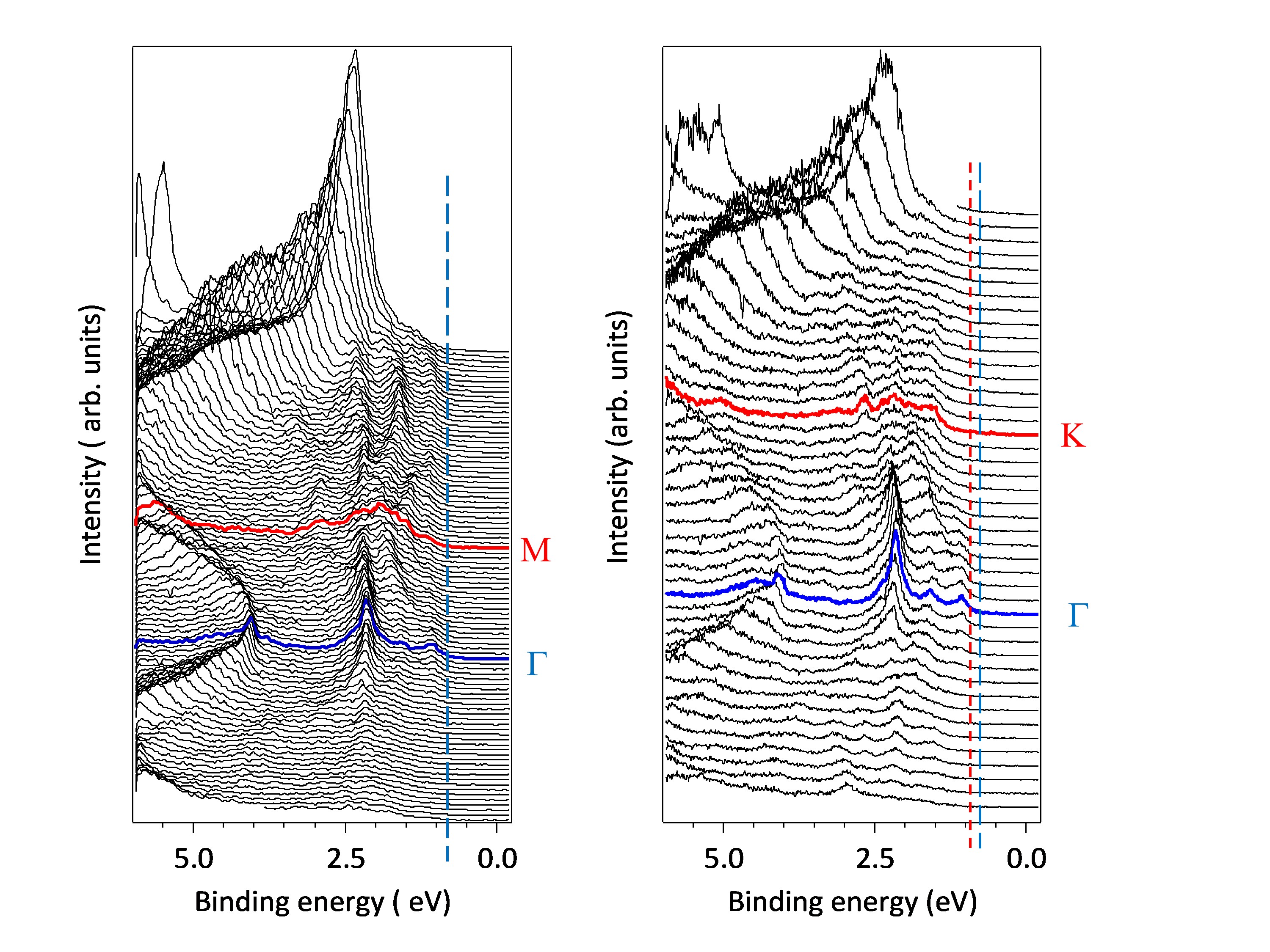}
  \caption{Photoemission energy distribution curves (EDCs) for monolayer ReSe$_2$ with momentum in the $\Gamma-M$ direction (left) and the $\Gamma-K$ direction (right); the EDC for $\Gamma$ is the lower, blue one. The vertical dashed lines for $\Gamma-K$ (right) at binding energies of $\sim0.75$ (long blue dashes) and $\sim0.88$~eV (short red dashes) are guides to the eye, marking estimates of the binding energy at which the photoemission signal rises above the baseline at the $\Gamma$ and $K$ points respectively; for the $\Gamma-M$ case, left hand side, the blue dashed line at a binding energy of $\sim0.8$~eV shows the position in energy of the nearly flat valence band edge.}
  \label{fgr:S4}
\end{figure}
\clearpage

\begin{figure}
  \includegraphics[width=\linewidth]{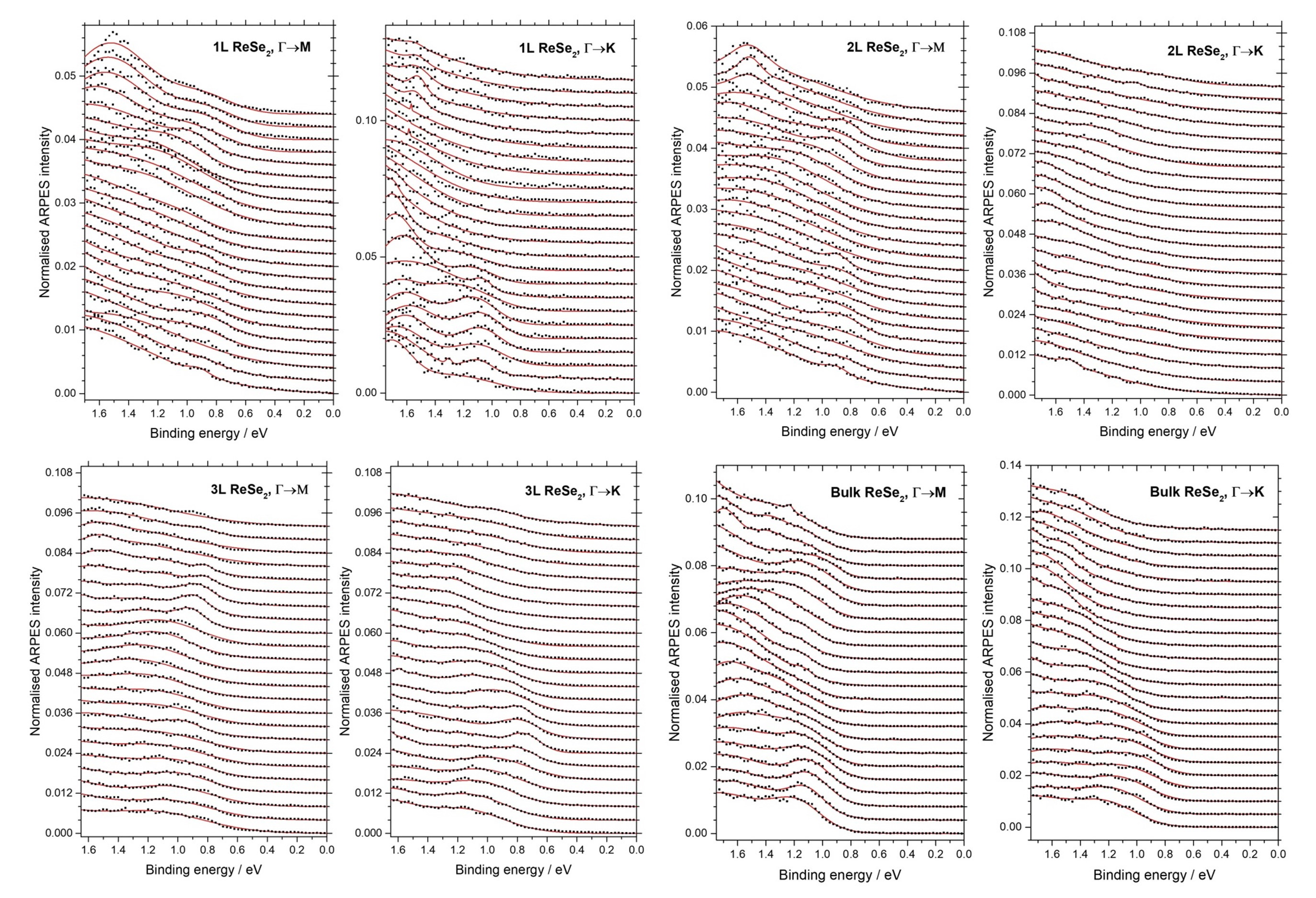}
  \caption{Photoemission energy distribution curves (EDCs; black dots) and fits to them (red lines) for 1L, 2L 3L and bulk  ReSe$_2$ in the directions $\Gamma-M$ and $\Gamma-K$ as indicated on the top right hand corner of each plot. The procedure for approximating the valence band edge energy is shown in the following figure, Fig.~\ref{fgr:S6}.}
  \label{fgr:S5}
\end{figure}
\clearpage

\begin{figure}
  \includegraphics[width=\linewidth]{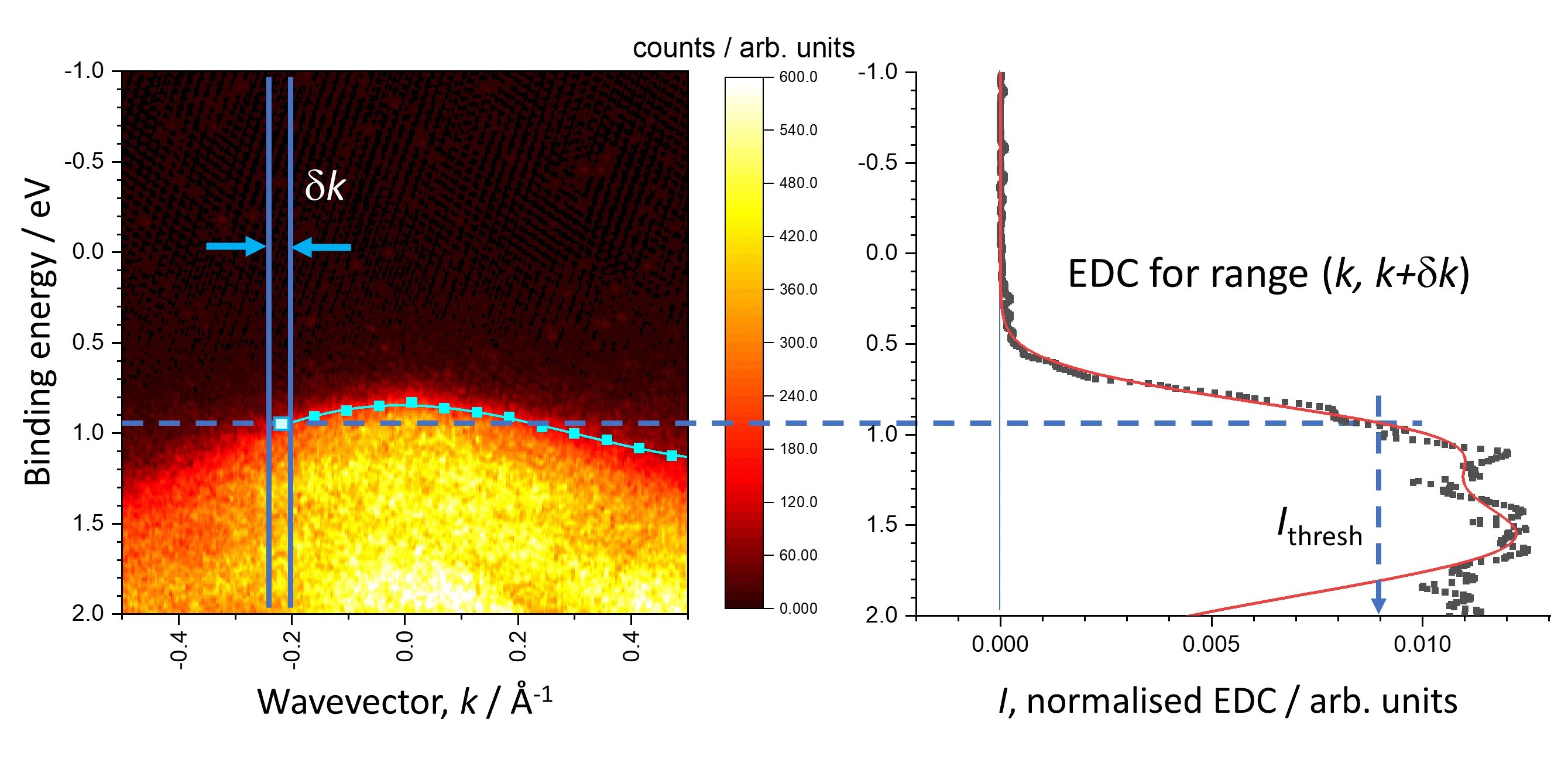}
  \caption{Fitting process used to extract the valence band edge position for each in-plane momentum $k$. Left: the counts for binding energies (BE) well above the valence band maximum (VBM) in the range $k$ to $k+\delta k$ were averaged over energy to obtain a baseline count; the baseline-corrected counts over this $k$ range were then summed over the whole energy range of interest to obtain a normalization value $N$. The normalized EDC for that $k$ was then obtained by summing the baseline-corrected counts at each energy over $k$ to $k+\delta k$ and dividing by $N$. Summing counts over a small range $\delta k \sim 0.01$~{\AA}$^{-1}$ reduced the noise in the EDC. Right: the EDC for that value of $k$ was fitted with Gaussians (two are shown here) to reproduce the shape of the normalized EDC within about 1~eV of the VBM. The binding energy at which a normalized threshold count $I_{\textrm{thresh}}$ close to the first Gaussian maximum is located was taken as an indicator of the band edge energy; note that, because of varying numbers of close-lying bands near the VBM, the peak positions of the Gaussians could \emph{not} be interpreted as the energy position of a particular band. $I_{\textrm{thresh}}$ was kept constant for all datasets for a given flake and was chosen to be as large as possible, so as close to the top of the highest band, without being unduly sensitive to the variations in fitting (the value shown here, 0.009, is typical, with a range of $\pm 0.001$). The sensitivity of the band edge positions to the chosen threshold gave the estimated uncertainties in bandwidth shown in Fig.~4(b) of the main text. The same, automated process was applied to all the experimental data for all flakes and all orientations (with a fixed choice of $\delta k$,  threshold, and initial Gaussian parameters); the band edge derived in this case is shown by the light blue points on the left hand panel. }
  \label{fgr:S6}
\end{figure}
\clearpage

\begin{figure}
\includegraphics[width=\linewidth]{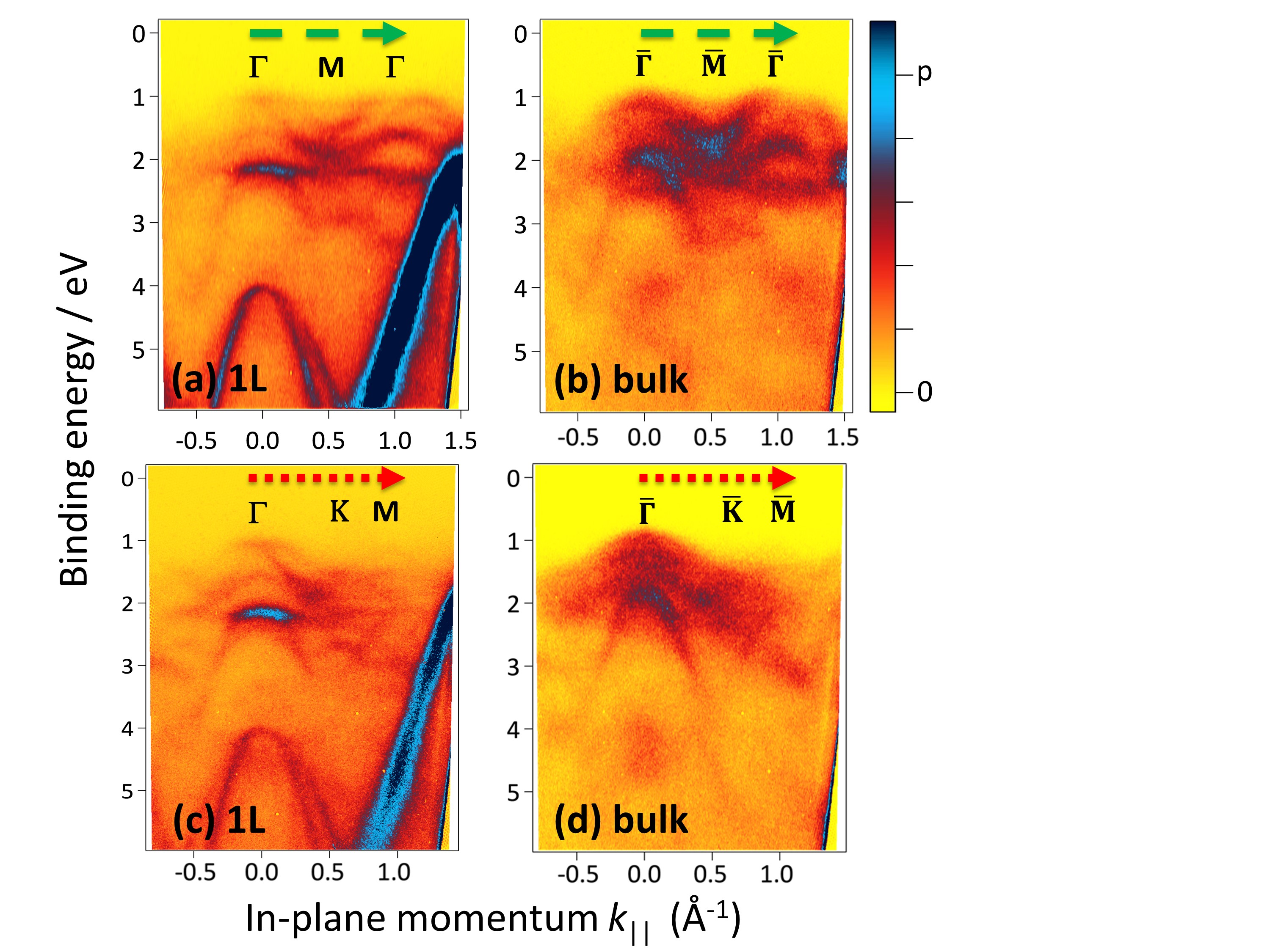}
\caption{Comparison of the measured valence band dispersions of a monolayer and a thick bulk flake with identical orientation. (a) and (b): monolayer and bulk dispersions in directions $\Gamma-M$; (c) and (d) monolayer and bulk dispersions in direction $\Gamma-K$. The color scale indicates counts from zero to $p$ where $p$ equals 1000, 800, 1400, 1000 for (a)-(d) respectively.}
\label{fgr:S7}
\end{figure}

\clearpage

\begin{figure}
  \includegraphics[width=\linewidth]{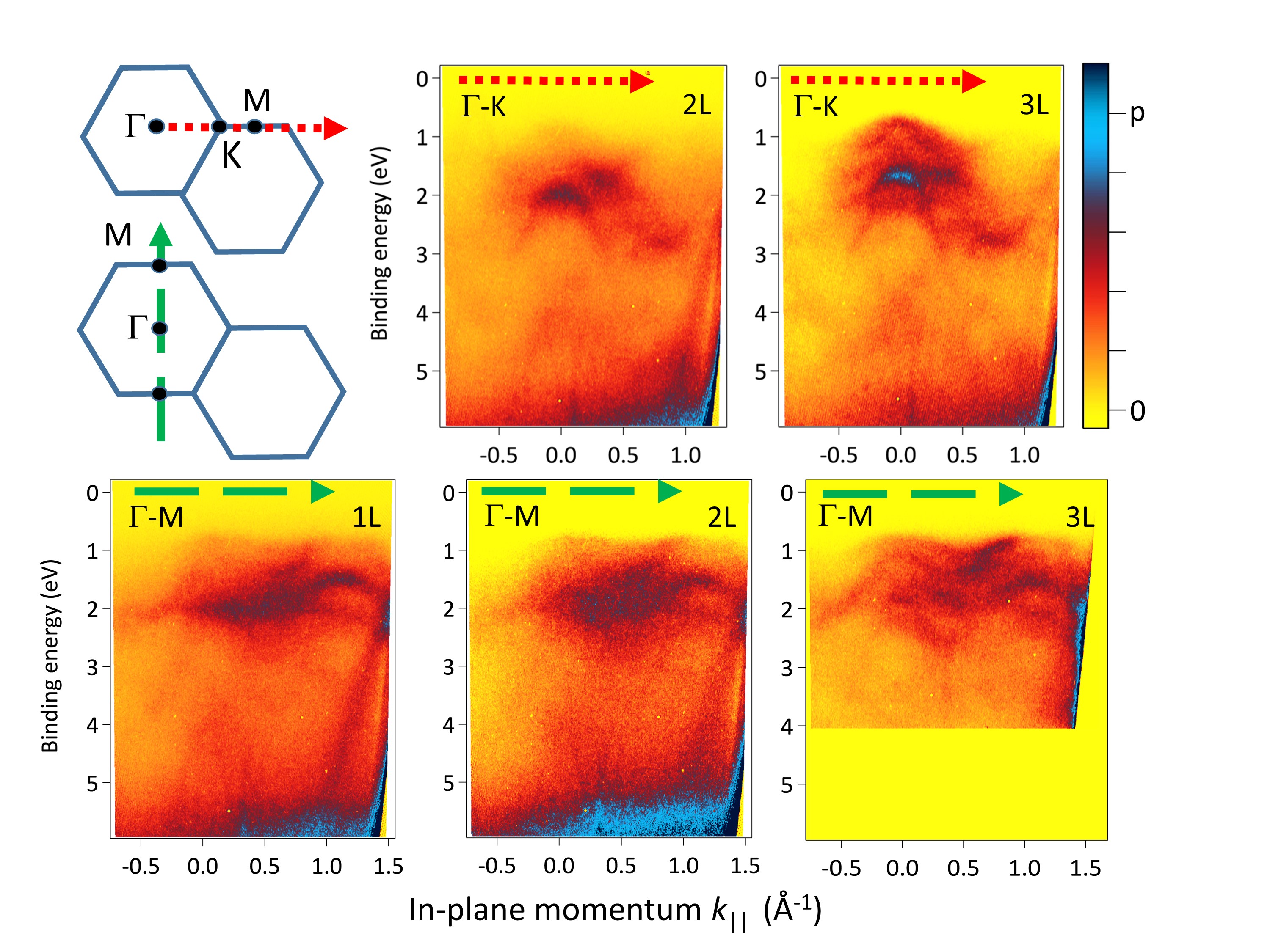}
  \caption{Comparison of the measured valence band dispersion along directions $\Gamma-K$ (dotted red arrow) and $\Gamma-M$ (dashed green arrow) for mono-, bi- and tri-layer flakes (1L, 2L and 3L), with schematic diagrams of the 2D Brillouin zone used to define the directions of the arrows. The data for $\Gamma-M$ is reproduced from the main paper for comparison. There is no data for $\Gamma-K$ for the 1L region of this sample, but $\Gamma-K$ data is shown for another 1L sample in the main paper. The color scale indicates counts from zero to $p$ where $p$ equals 3200, 3200 for the top row and 3000, 1600, 1600 for the bottom row.}
  \label{fgr:S8}
\end{figure}
\clearpage

\begin{figure}
  \includegraphics[width=\linewidth]{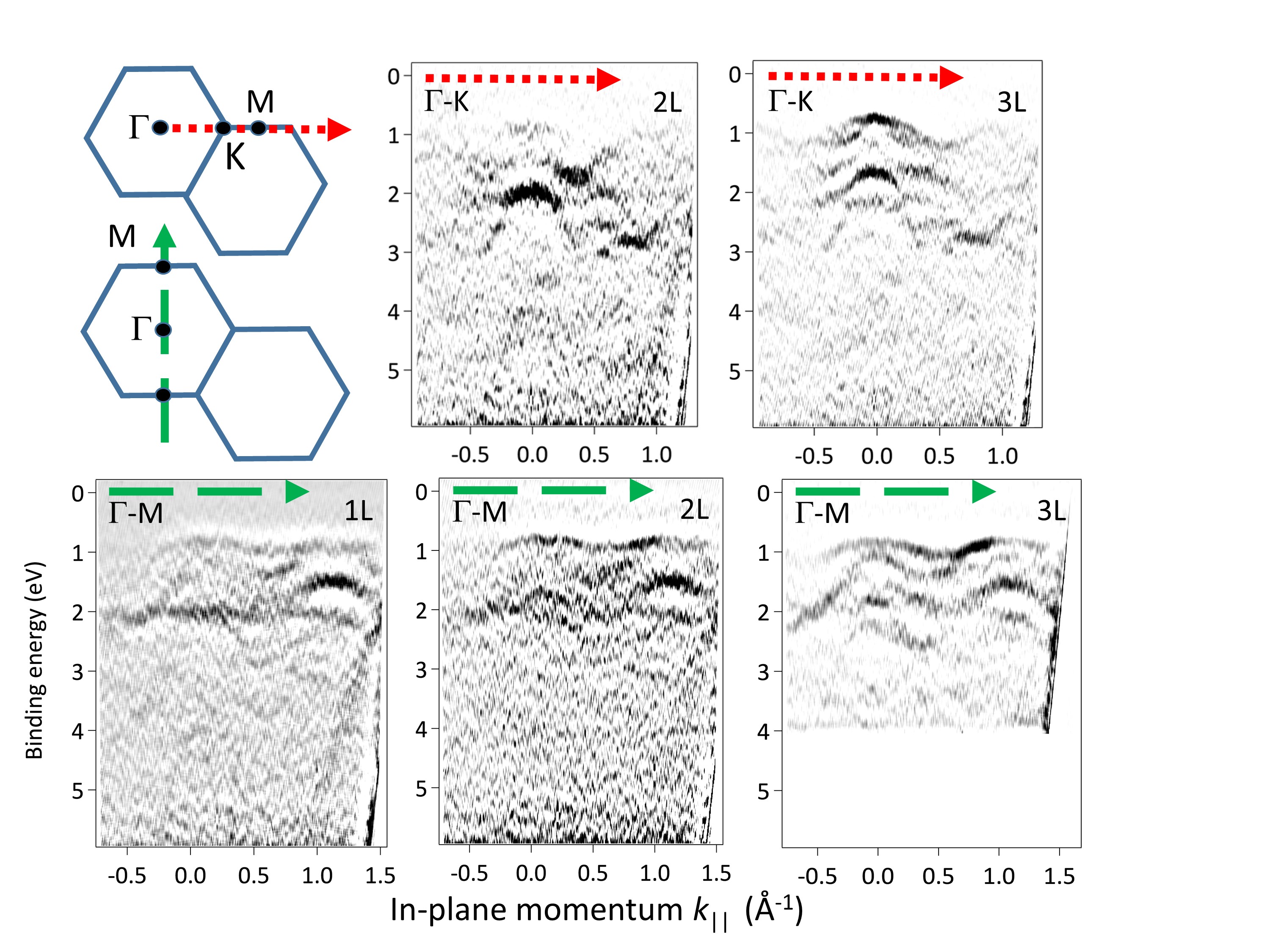}
  \caption{Comparison of the twice-differentiated valence band dispersion along directions $\Gamma-K$ (dotted red arrow) and $\Gamma-M$ (dashed green arrow) for mono-, bi- and tri-layer flakes (1L, 2L and 3L); the raw data is shown in Fig.~\ref{fgr:S8}.}
  \label{fgr:S9}
\end{figure}
\clearpage

\section{Results of calculations.}

\begin{figure}[h]
  \includegraphics[width=\linewidth]{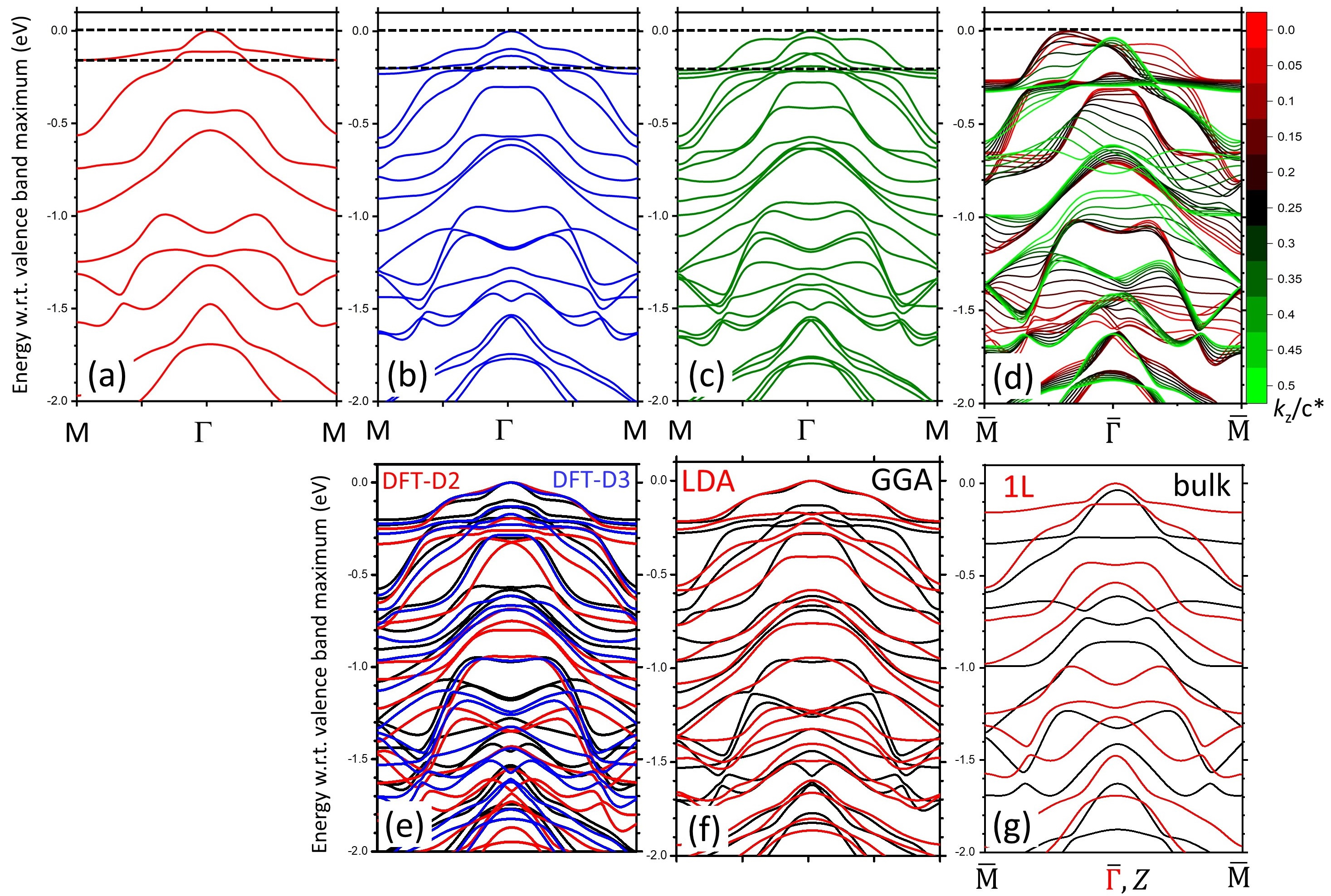}
  \caption{Calculated $\Gamma-M$ dispersions for (a) 1L, (b) 2L, (c) 3L and (d) bulk ReSe$_2$. In (d), dispersions for a set of equally-spaced slices through the Brillouin zone normal to $c^*$ are shown, representing the bulk continuum; panel (d) is symmetrical about $\Gamma$ if slices with $-0.5 c^*<k_z<0.5 c^*$ are included, since the 2D projection has inversion symmetry but, for clarity, we only plot dispersions in the range $0<k_z<0.5 c^*$ (red to green). The dashed black lines show the energies of the extrema of the top valence band in each case. Panel (e) compares the 2L dispersions including Grimme DFT-D2 and DFT-D3 inter-layer van der Waals (vdW) corrections \cite{Grimme2010}. In (a)-(d), scalar relativistic GGA projector-augmented wave functionals were used, while (f) compares LDA and GGA PAW (no vdW corrections). Good agreement with the layer spacing (6.275~{\AA}) for bulk \cite{LAMFERS199634} is given by the GGA+D3 (6.288~{\AA}) and fully-relativistic LDA (6.359~{\AA}) calculations. Panel (g) compares the 1L dispersions to the bulk dispersion passing through $Z$ ($k_z=c^*/2$). }
  \label{fgr:S10}
\end{figure}
\clearpage

\begin{figure}
  \includegraphics[width=\linewidth]{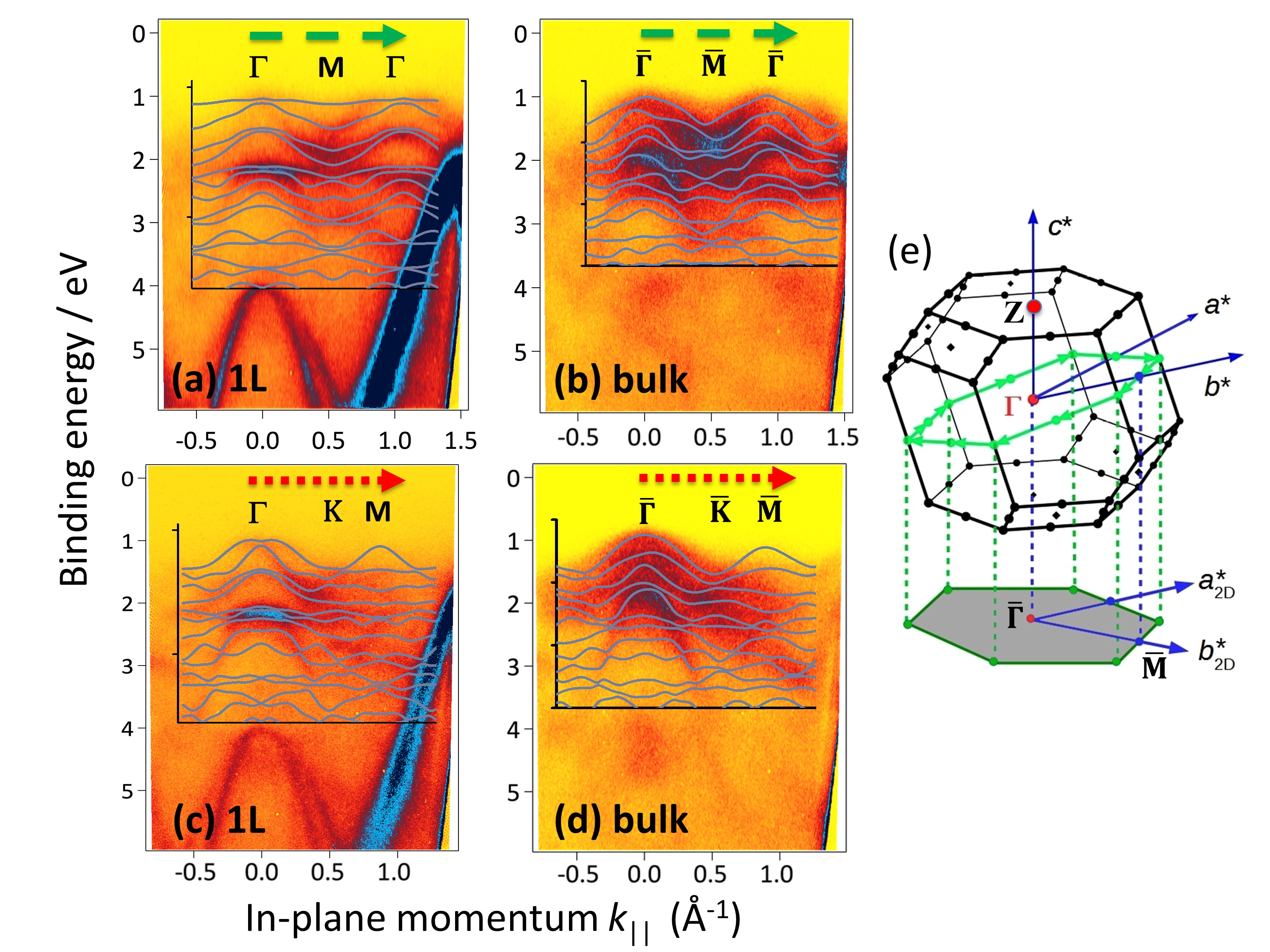}
  \caption{Comparison of calculated and experimental valence band dispersions of ReSe$_2$ in the (a,b) $\Gamma-M$ and (c,d) orthogonal $\Gamma-K$ directions for (a,c) monolayer and (b,d) a bulk-like thick flake. The experimental orientations of the bulk and monolayer flakes were identical (they were adjacent on the substrate after exfoliation). (e) shows the bulk 3D and monolayer 2D Brillouin zones with the projected $\bar K$ and $\bar M$ directions indicated in the latter. The calculations for the bulk dispersion took into account the dependence of $k_z$ on the in-plane momentum using an inner potential of 19~eV \cite{Hart2017,Eickholt2018} and the excitation photon energy of 100~eV. At this energy, we expect to probe the VB near the $Z$ point \cite{Hart2017,Kim2019} indicated in (e). Here, fully-relativistic projector augmented wave (PAW) pseudopotentials were used in the LDA approximation in both 1L and bulk cases. The data and color scales in this figure are the same as in Fig.~\ref{fgr:S7}.}
  \label{fgr:S11}
\end{figure}
\clearpage

\begin{figure}
  \includegraphics[width=\linewidth]{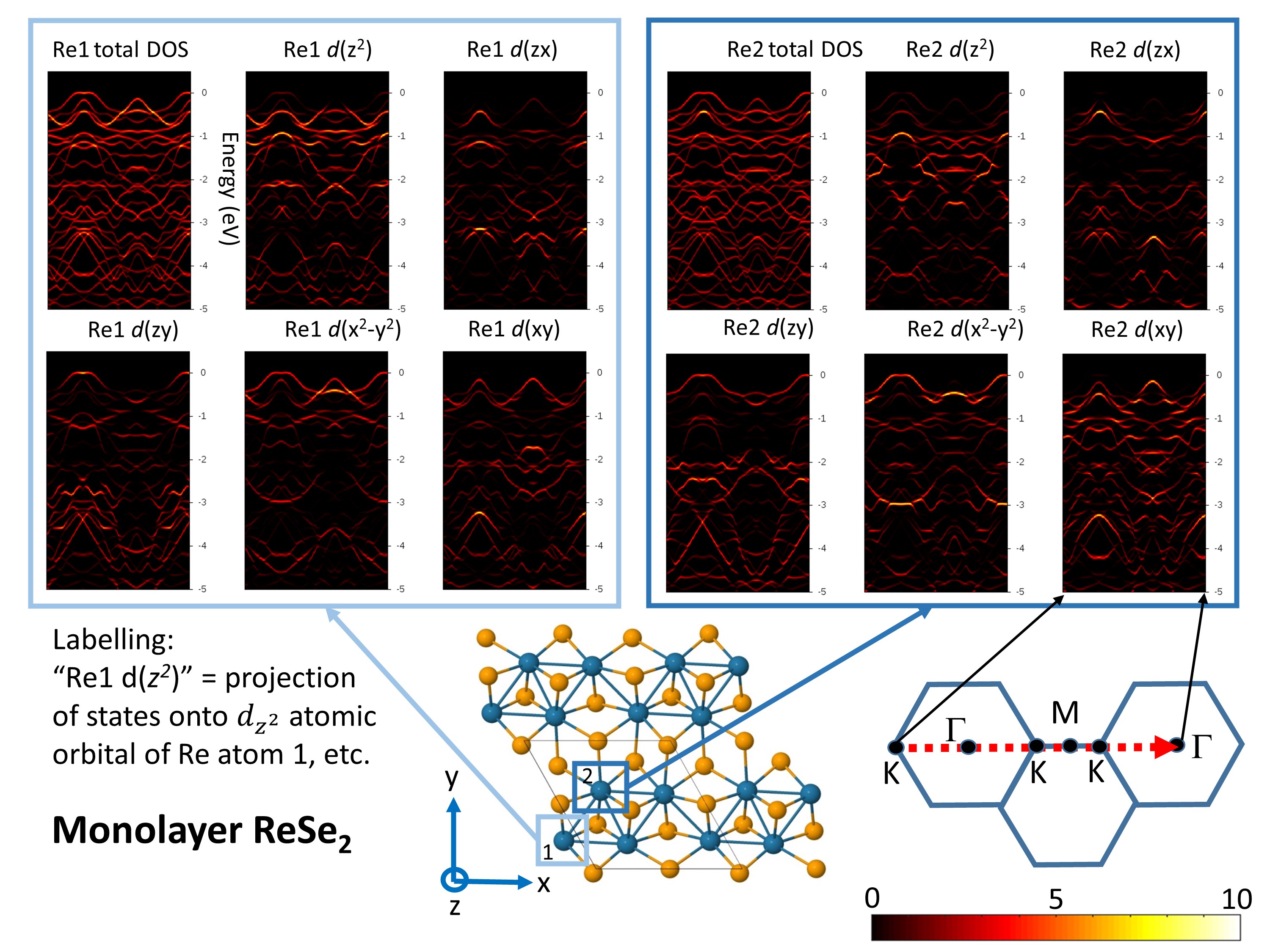}
  \caption{Calculated valence band dispersion of monolayer ReSe$_2$ projected onto atomic orbitals of Re (calculations here used the scalar-relativistic GGA approximation). The top of the valence band is defined as the zero of energy: momentum is along the direction $\Gamma-K$ as indicated with respect to the 2D Brillouin zone (dotted red arrow). Left light blue box: projections onto $d$ orbitals of Re atom 1; Right dark blue box: projections onto $d$ orbitals of Re atom 2. The other two Re atoms are related to these two by inversion. The color scale in units of states per eV applying to this and all following figures is shown, bottom right.}
  \label{fgr:S12}
\end{figure}

Figure~\ref{fgr:S12} shows that, in comparison to the projected states of the Se atoms, the Re atoms at the two non-equivalent sites make similar contributions to the band structure. The compositions of the top valence band near $\Gamma$ and near $K$ are a key question; unlike many better-known transition metal dichalcogenides, the contribution of Re $d_{z^2}$ at the top of the valence band is small compared to that of Re $d_{x^2-y^2}$ across the whole of the dispersion.
\clearpage

\begin{figure}
  \includegraphics[width=\linewidth]{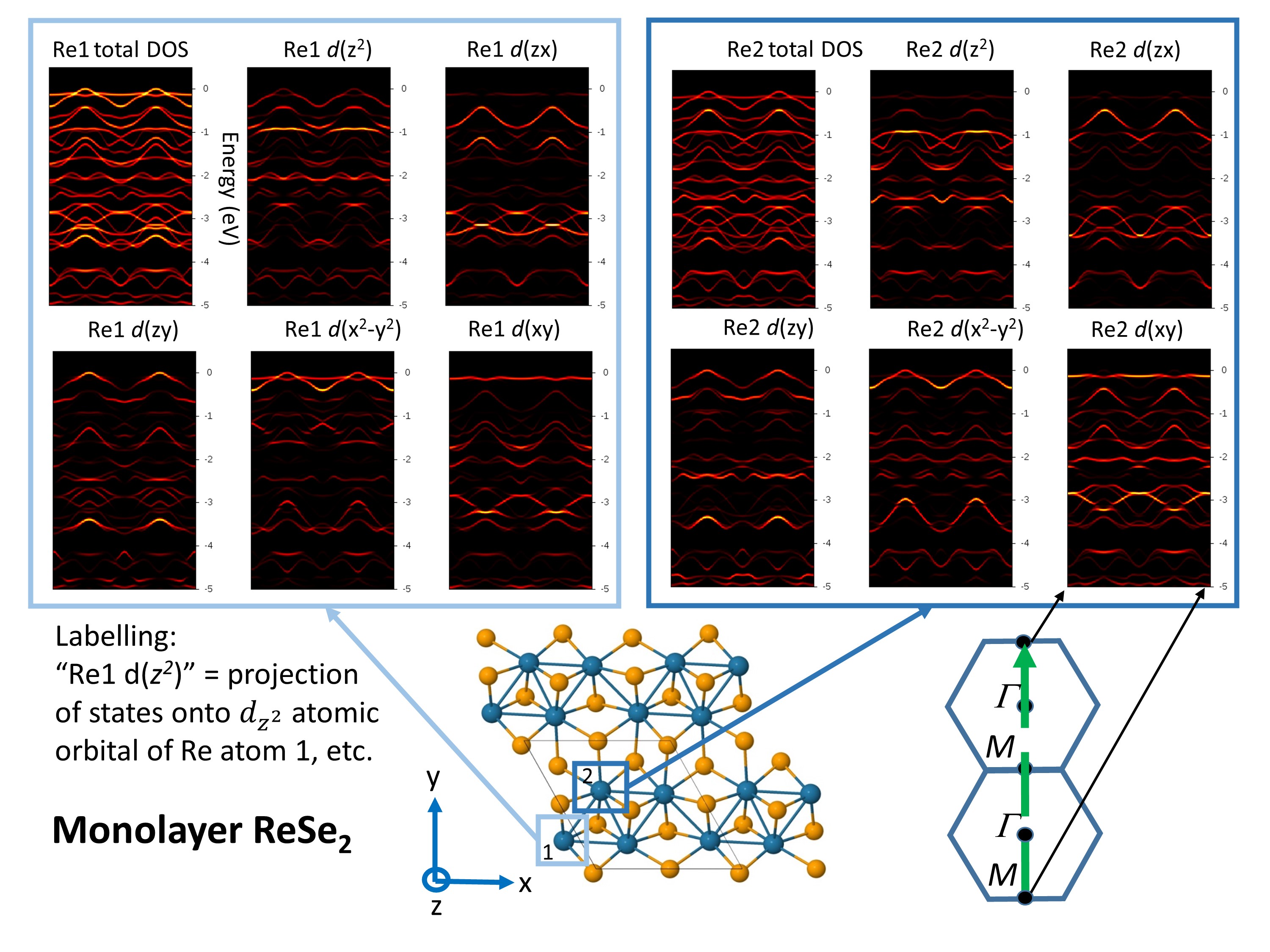}
  \caption{Calculated valence band dispersion of monolayer ReSe$_2$ projected onto atomic orbitals of Re (calculations here used the scalar-relativistic GGA approximation). The momentum is along the direction $\Gamma-M$ as indicated with respect to the 2D Brillouin zone (dashed green arrow).}
  \label{fgr:S13}
\end{figure}

Figure~\ref{fgr:S13} demonstrates that the flat band forming the top of the valence band in the $\Gamma-M$ direction has a strong contribution from the Re $d_{x^2-y^2}$ and $d_{xy}$ states; as in Figure~\ref{fgr:S12}, the contribution of $d_{z^2}$ at the top of the valence band is again much smaller. On the other hand, the $d_{z^2}$ states of both Re atoms make a significant contribution near $\Gamma$ at $\sim1$~eV below the top of the valence band; by reference to Fig.~S15, one can see that these are hybridised principally with Se $p_z$ orbitals. This is the origin of one prominent feature in the experimental ARPES data.
\clearpage

\begin{figure}
  \includegraphics[width=\linewidth]{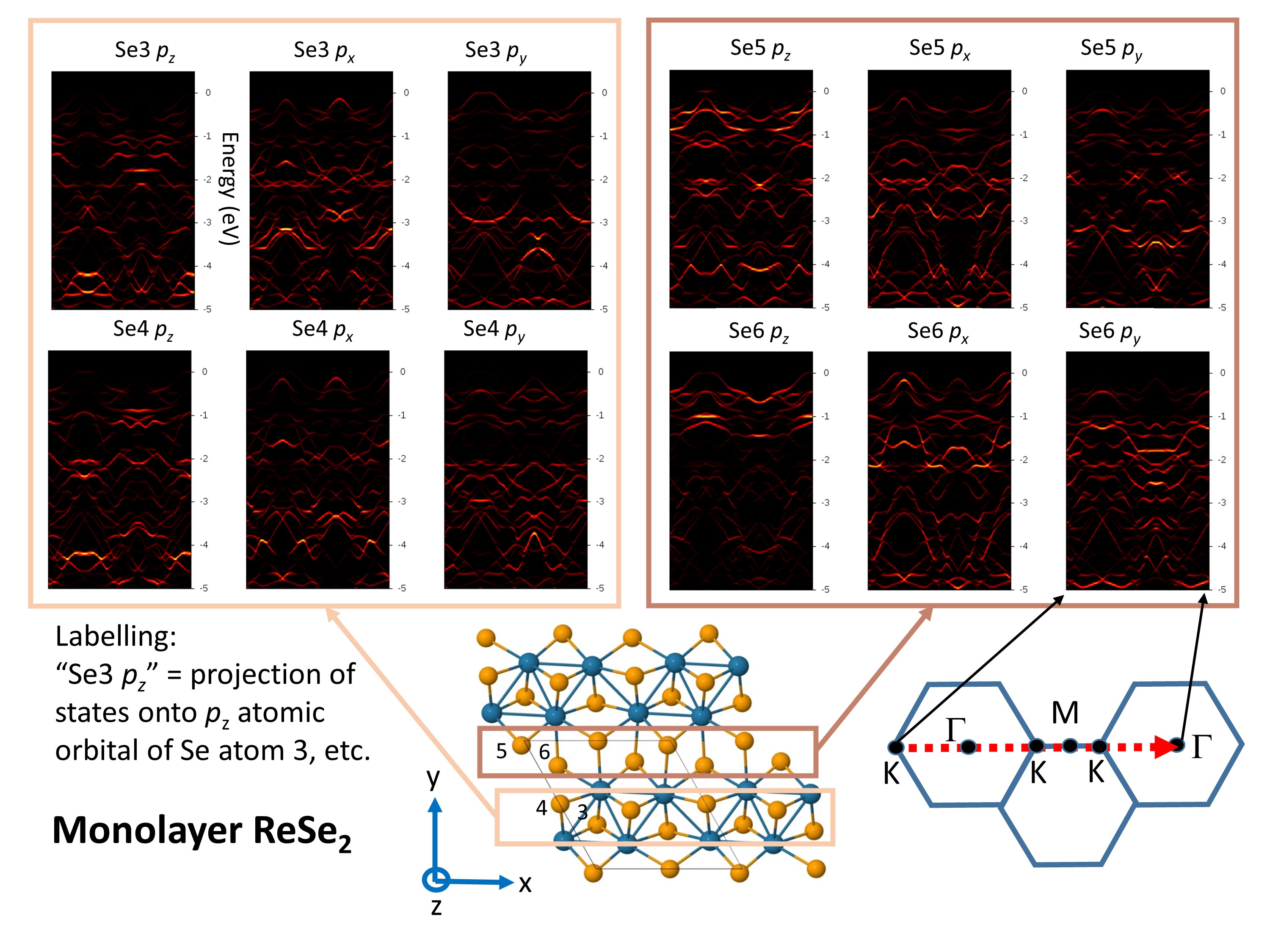}
  \caption{Calculated valence band dispersion of monolayer ReSe$_2$ along the direction $\Gamma-K$ projected onto atomic orbitals of Se (calculations here used the scalar-relativistic GGA approximation). Left hand panel (orange box): $p_z, p_x, p_y$ states (left to right) of selenium atoms 3 and 4; Right hand panel (brown box): projections onto the same states for selenium atoms 5 and 6. The other selenium atoms in the unit cell are related to these four by inversion symmetry. The range of the momentum slice is indicated by reference to the 2D Brillouin zone.}
  \label{fgr:S14}
\end{figure}

\begin{figure}
  \includegraphics[width=\linewidth]{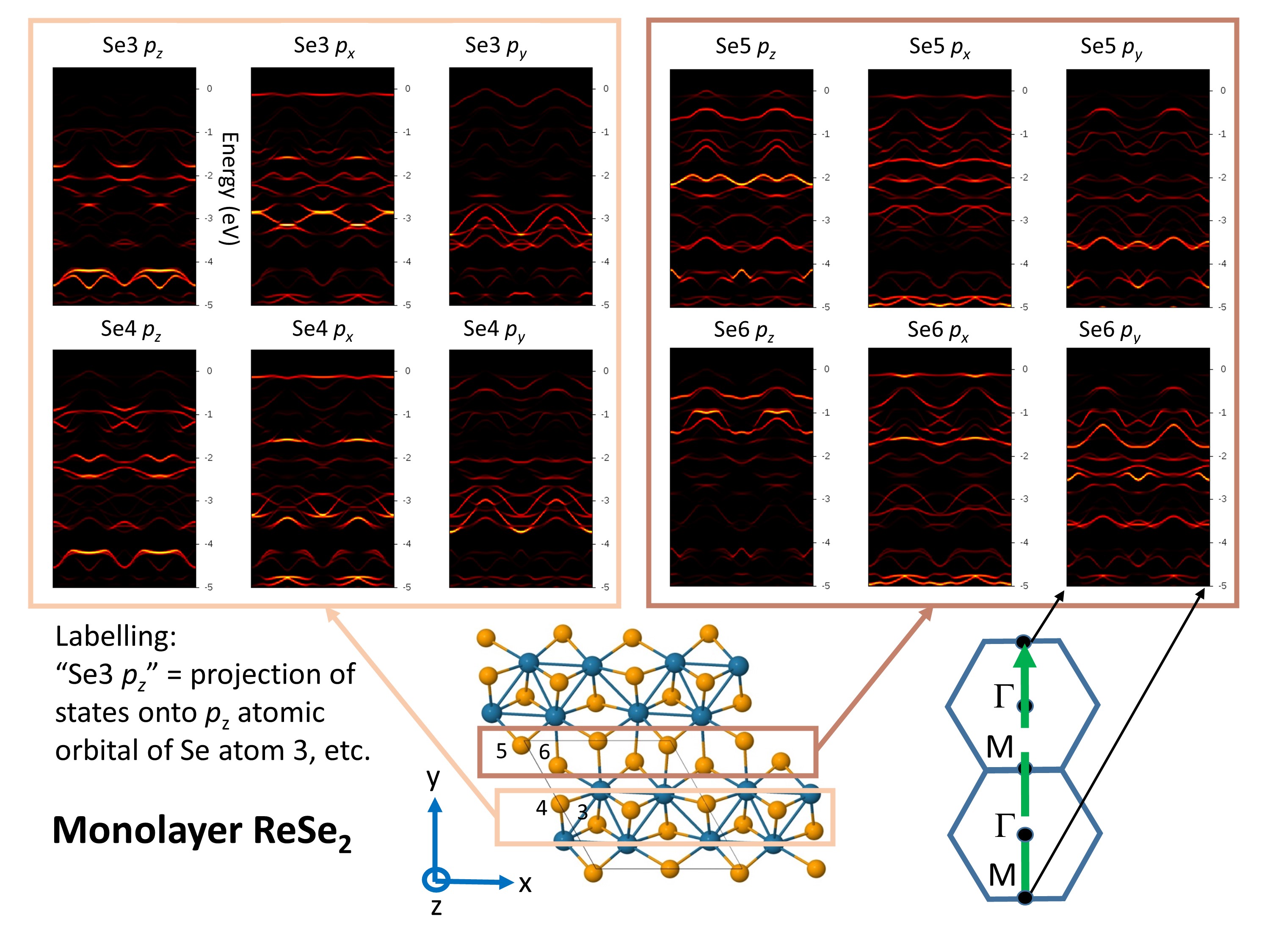}
  \caption{Calculated valence band dispersion of monolayer ReSe$_2$ projected onto atomic orbitals of Se (calculations here used the scalar-relativistic GGA approximation). The momentum is along the orthogonal direction $\Gamma-M$ as indicated with respect to the 2D Brillouin zone (dashed green arrow).}
  \label{fgr:S15}
\end{figure}

The grouping of the dispersions in Fig.~\ref{fgr:S14} highlights the fact that the Se atoms can be assigned to two groups according to their contributions to the valence band structure. The Se atoms (3 and 4) above and below the Re chains have projections which resemble each other much more closely than they do those of atoms 5 and 6 (see, e.g, the $p_z$ contributions near the top of the valence band); see also Fig.~\ref{fgr:S15} for the direction $\Gamma-M$.

\clearpage

In Figure~\ref{fgr:S15}, the very flat band in the $\Gamma-M$ direction is clearly associated with the $p_x$ orbitals of all four Se atoms, especially Se3 $p_x$ and Se4 $p_x$. Once again, the difference between the contributions of Se atoms 3 and 4 to those of atoms 5 and 6 is clear (see, for example, the dominant contributions of the $p_z$ and $p_y$ states, which are very different).
\clearpage

\begin{figure}
  \includegraphics[width=\linewidth]{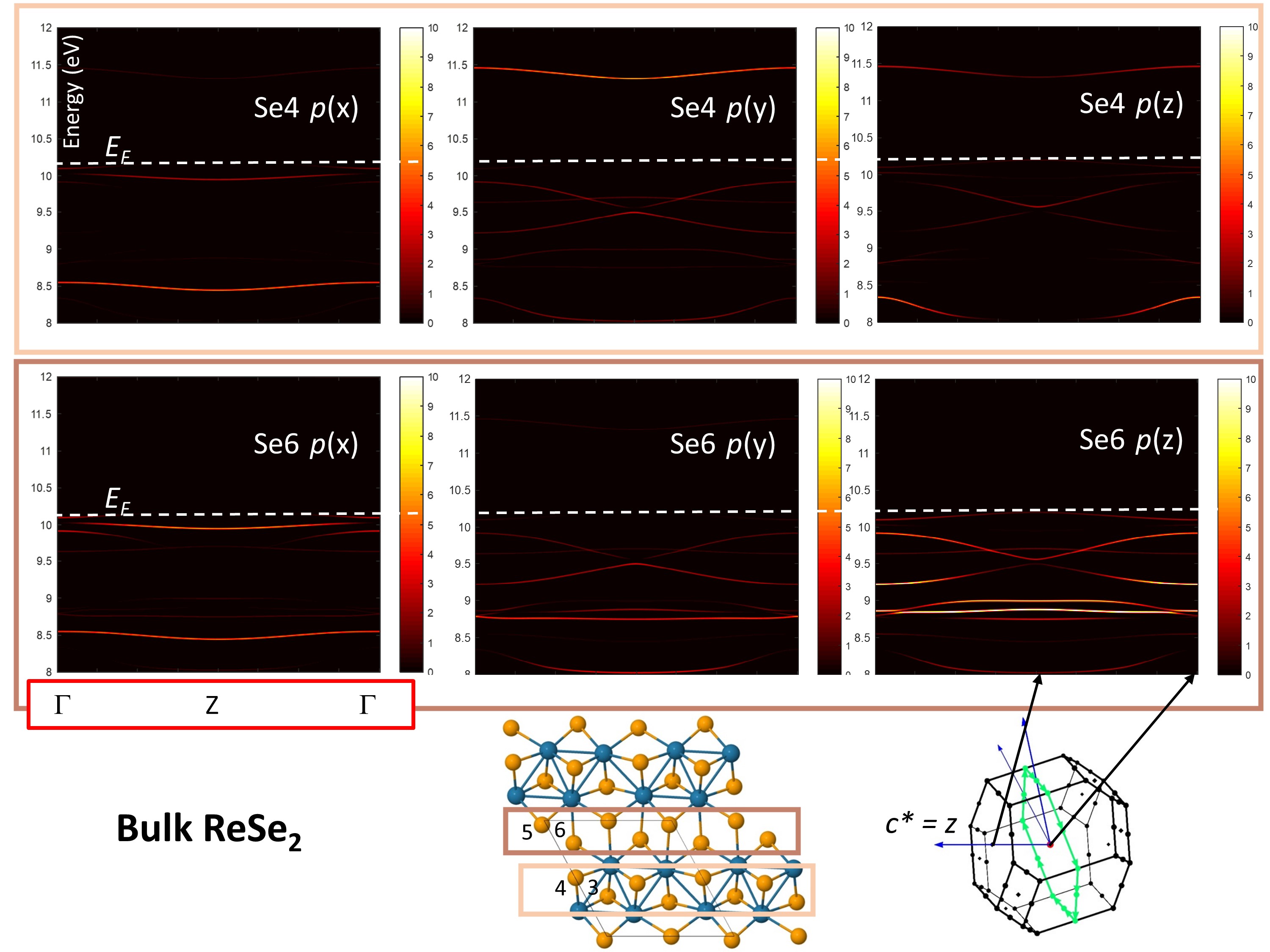}
  \caption{Calculated dispersion of the valence band of bulk ReSe$_2$ projected onto atomic orbitals (calculations here used the scalar-relativistic GGA approximation and forces were relaxed below $\sim$0.1 eV${\AA}^{-1}$.) for the direction $\Gamma-Z$, perpendicular to the layer planes. Examples of the two types of Se atom are considered (top orange box: atom 4; bottom brown box: atom 6); the other two atoms (3 and 5) give similar results. The energy of the top of the valence band is indicated by the white dashed lines.}
  \label{fgr:S16}
\end{figure}

Similar to other DFT-based calculations and experimental data for bulk ReSe$_2$, \cite{Hart2017, Arora2017, Echeverry2018, Eickholt2018, Kim2019} we find that the $Z$ point is highest in energy along this path in reciprocal space, as shown in Fig.~\ref{fgr:S16}. The chalcogen contribution to the uppermost band switches between dominant $p_z$ character at $Z$ to $p_x$ at $\Gamma$ and both lie very close in energy. In the next figure, \ref{fgr:S17}, the contributions of the Re atoms to the same bands are analyzed.

\clearpage

\begin{figure}
  \includegraphics[width=\linewidth]{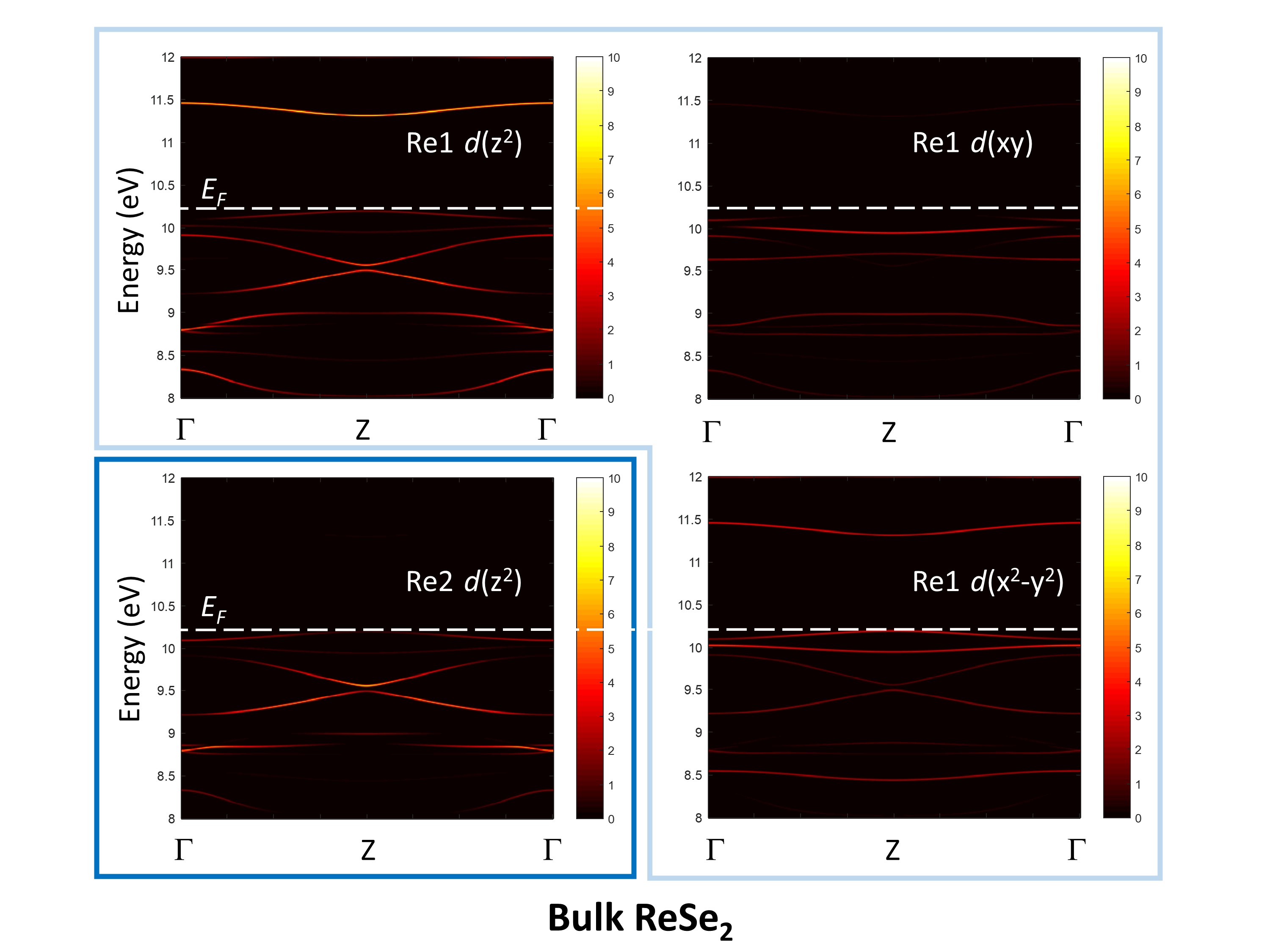}
  \caption{Calculated dispersion of the valence band of bulk ReSe$_2$ projected onto atomic orbitals (neglecting spin-orbit coupling) for the direction $\Gamma-Z$, perpendicular to the layer planes, as for Fig.~\ref{fgr:S16}. The two types of Re atom are considered (top light blue box: atom 1; bottom dark blue box: atom 2). The energy of the top of the valence band is indicated by the white dashed lines.}
  \label{fgr:S17}
\end{figure}

Fig.~\ref{fgr:S17} shows that the contributions of the Re1 $d_{z^2}$ and $d_{x^2-y^2}$ states to the uppermost valence band are dominant, and the changes in composition of the top band on moving from $Z$ to $\Gamma$ are only slight.
\clearpage

\bibliography{ReSe2_Refs}

\closesupplement